\documentclass[sigconf]{acmart}

\PassOptionsToPackage{table}{xcolor}
\setcopyright{none} 

\renewcommand\footnotetextcopyrightpermission[1]{}
\acmConference{CCS 2022}{Los Angeles, US}

\usepackage{macros}
\usepackage{graphicx}
\usepackage[normalem]{ulem}

\usepackage{pifont}
\usepackage{booktabs} 
\usepackage{amsmath}
\usepackage{wasysym}
\usepackage{subfigure}
\usepackage{colortbl}

\usepackage{algorithmicx}
\usepackage{algpseudocode}
\usepackage[boxed]{algorithm}
\usepackage{macros}
\usepackage{ioa_code}
\usepackage{graphicx}
\usepackage{framed}
\usepackage{collcell}
\usepackage{hhline}
\usepackage{pgf}
\usepackage{multirow}

\usepackage{collcell}
\usepackage{hhline}
\usepackage{pgf}
\usepackage{multirow}
\def\colorModel{hsb} 
\newcommand\ColCell[1]{
  \pgfmathparse{#1<100?1:0}  
    \ifnum\pgfmathresult=0\relax\color{white}\fi
  \pgfmathsetmacro\compA{1}      
  \pgfmathsetmacro\compB{#1/332}   
  \pgfmathsetmacro\compC{1}      
  \edef\x{\noexpand\centering\noexpand\cellcolor[\colorModel]{\compA,\compB,\compC}}\x #1
  } 
\newcolumntype{E}{>{\collectcell\ColCell}m{0.7cm}<{\endcollectcell}}  

\algnewcommand\algorithmicforeach{\textbf{for each}}
\algdef{S}[FOR]{ForEach}[1]{\algorithmicforeach\ #1\ \algorithmicdo}

\usepackage{xspace}

\def\colorModel{hsb}

\newtheorem{theorem}{Theorem}
\newtheorem{lemma}{Lemma}
\newtheorem{definition}{Definition}

\PassOptionsToPackage{hyphens}{url}
\usepackage{hyperref}

\usepackage{url}
\makeatletter
\g@addto@macro{\UrlBreaks}{\UrlOrds}
\makeatother

\hypersetup{
	colorlinks=true,
	linkcolor=blue,
	citecolor=red,
	urlcolor=magenta,
}
\let\labelindent\relax
\usepackage{enumitem}
\usepackage{xspace}

\newcommand{\vincent}[1]{{\color{red}{Vincent: #1}}}
\newcommand{\deepal}[1]{{\color{blue}{Deepal: #1}}}
\newcommand{\cref}[1]{{\S\ref{#1}}}

\newcommand{\remove}[1]{}

\newenvironment{smallitem}{
\begin{itemize}[%
leftmargin=10pt,
labelsep=5pt,
rightmargin=3pt,
labelwidth=0pt,
itemindent=2pt,
listparindent=5pt,
topsep=4pt plus 2pt minus 4pt,
partopsep=4pt,
itemsep=6pt,
parsep=5pt
]
    \setlength{\parskip}{-1pt}
}{\end{itemize}}

\let\emptyset\varnothing

\usepackage{graphicx}

\usepackage{framed}

\usepackage{listings}
\usepackage{multicol}
\usepackage{color}
\usepackage{booktabs}
\definecolor{black}{RGB}{0,0,0}
\definecolor{gray}{RGB}{102,102,102}        
\definecolor{function}{RGB}{0,102,153}      
\definecolor{lightgreen}{RGB}{102,153,0}    
\definecolor{lightlightgreen}{RGB}{152,193,50}    
\definecolor{bluegreen}{RGB}{51,153,126}    
\definecolor{magenta}{RGB}{217,74,122}  
\definecolor{orange}{RGB}{226,102,26}       
\definecolor{purple}{RGB}{125,71,147}       
\definecolor{green}{RGB}{113,138,98}        
\definecolor{tomato}{RGB}{255,99,71}  
\definecolor{lightred}{RGB}{255,160,131}  
\usepackage[T1]{fontenc}
\usepackage[scaled]{beramono}
\usepackage{microtype}
\usepackage{float}
\lstdefinelanguage{parameterized}{
  firstnumber=1,
  xleftmargin=1.5em,
  numberstyle=\tiny\color{black},
  tabsize=2,
  numbers=left,
  morekeywords = [3]{require,while,if,then,else,do,done,wait,until,end,for,return,returns,upon,from,to,is,in},
  morekeywords = [4]{pragma,function,contract,new,true,false,null,and,or},
  morekeywords = [5]{bench,unlockAccount,createBatch,start_new_consensus,peek,
  propose,execute_transaction,HttpProvider,parse,on,catch,readFileSync,sendSignedTransaction,
  consensus_propose,mvc_propose,break,decide,poll,broadcast,deliver,add,greet,Hello,
  execute,sendTransaction,updateBlockState,executeTx,persist,deserialize},
  morekeywords = [6]{var,public,bytes32,bool,byte,+,=,:=,.,;,,,-,!,=,~,>,<,==,solidity},
  morekeywords = [7]{reliable_broadcast,binary_consensus,consensus},
  keywordstyle = [3]\color{bluegreen},
  keywordstyle = [4]\color{lightgreen},
  keywordstyle = [5]\color{magenta},
  keywordstyle = [6]\color{orange},
  keywordstyle = [7]\color{purple},
  sensitive = true,
  morecomment = [l][\color{gray}]{//},
  morecomment = [s][\color{gray}]{/*}{*/},
  morecomment = [s][\color{gray}]{/**}{*/},
  morestring = [b][\color{purple}]",
  morestring = [b][\color{purple}]',
  literate=
  {á}{{\'a}}1 {é}{{\'e}}1 {í}{{\'i}}1 {ó}{{\'o}}1 {ú}{{\'u}}1
  {Á}{{\'A}}1 {É}{{\'E}}1 {Í}{{\'I}}1 {Ó}{{\'O}}1 {Ú}{{\'U}}1
  {à}{{\`a}}1 {è}{{\`e}}1 {ì}{{\`i}}1 {ò}{{\`o}}1 {ù}{{\`u}}1
  {À}{{\`A}}1 {È}{{\'E}}1 {Ì}{{\`I}}1 {Ò}{{\`O}}1 {Ù}{{\`U}}1
  {ä}{{\"a}}1 {ë}{{\"e}}1 {ï}{{\"i}}1 {ö}{{\"o}}1 {ü}{{\"u}}1
  {Ä}{{\"A}}1 {Ë}{{\"E}}1 {Ï}{{\"I}}1 {Ö}{{\"O}}1 {Ü}{{\"U}}1
  {â}{{\^a}}1 {ê}{{\^e}}1 {î}{{\^i}}1 {ô}{{\^o}}1 {û}{{\^u}}1
  {Â}{{\^A}}1 {Ê}{{\^E}}1 {Î}{{\^I}}1 {Ô}{{\^O}}1 {Û}{{\^U}}1
  {Ã}{{\~A}}1 {ã}{{\~a}}1 {Õ}{{\~O}}1 {õ}{{\~o}}1
  {½}{{\oe}}1 {¼}{{\OE}}1 {æ}{{\ae}}1 {Æ}{{\AE}}1 {ß}{{\ss}}1
  {?}{{\H{u}}}1 {?}{{\H{U}}}1 {?}{{\H{o}}}1 {?}{{\H{O}}}1
  {ç}{{\c c}}1 {Ç}{{\c C}}1 {ø}{{\o}}1 {å}{{\r a}}1 {Å}{{\r A}}1
  {¤}{{\euro}}1 {£}{{\pounds}}1 {«}{{\guillemotleft}}1
  {»}{{\guillemotright}}1 {ñ}{{\~n}}1 {Ñ}{{\~N}}1 {¿}{{?`}}1
}
\newcommand*\LSTfont{\Small\ttfamily\SetTracking{encoding=*}{-60}\lsstyle}
\newcommand\TextSize{\fontsize{8.5}{9.5}\selectfont}
\newcommand*\ttt{\TextSize\ttfamily\SetTracking{encoding=*}{-60}\lsstyle}

\definecolor{verylightgray}{rgb}{.97,.97,.97}

\lstdefinelanguage{Solidity}{
        tabsize=2,
        numbers=left,
        stepnumber=1,
        showstringspaces=false,
	keywords=[1]{anonymous, assembly, assert, balance, break, call, callcode, case, catch, class, constant, continue, constructor, contract, debugger, default, delegatecall, delete, do, else, emit, event, experimental, export, external, false, finally, for, function, gas, if, implements, import, in, indexed, instanceof, interface, internal, is, length, library, log0, log1, log2, log3, log4, memory, modifier, new, payable, pragma, private, protected, public, pure, push, require, return, returns, revert, selfdestruct, send, solidity, storage, struct, suicide, super, switch, then, this, throw, transfer, true, try, typeof, using, value, view, while, with, addmod, ecrecover, keccak256, mulmod, ripemd160, sha256, sha3}, 
	keywordstyle=[1]\color{lightgreen}\bfseries,
	keywords=[2]{address, bool, byte, bytes, bytes1, bytes2, bytes3, bytes4, bytes5, bytes6, bytes7, bytes8, bytes9, bytes10, bytes11, bytes12, bytes13, bytes14, bytes15, bytes16, bytes17, bytes18, bytes19, bytes20, bytes21, bytes22, bytes23, bytes24, bytes25, bytes26, bytes27, bytes28, bytes29, bytes30, bytes31, bytes32, enum, int, int8, int16, int24, int32, int40, int48, int56, int64, int72, int80, int88, int96, int104, int112, int120, int128, int136, int144, int152, int160, int168, int176, int184, int192, int200, int208, int216, int224, int232, int240, int248, int256, mapping, string, uint, uint8, uint16, uint24, uint32, uint40, uint48, uint56, uint64, uint72, uint80, uint88, uint96, uint104, uint112, uint120, uint128, uint136, uint144, uint152, uint160, uint168, uint176, uint184, uint192, uint200, uint208, uint216, uint224, uint232, uint240, uint248, uint256, var, void, ether, finney, szabo, wei, days, hours, minutes, seconds, weeks, years},	
	keywordstyle=[2]\color{bluegreen}\bfseries,
	keywords=[3]{block, blockhash, coinbase, difficulty, gaslimit, number, timestamp, msg, data, gas, sender, sig, value, now, tx, gasprice, origin},	
	keywordstyle=[3]\color{function}\bfseries,
	identifierstyle=\color{black},
	sensitive=false,
	comment=[l]{//},
	morecomment=[s]{/*}{*/},
	commentstyle=\color{gray}\ttfamily,
	stringstyle=\color{purple}\ttfamily,
	morestring=[b]',
	morestring=[b]"
}

\makeatletter
\newcommand\footnoteref[1]{\protected@xdef\@thefnmark{\ref{#1}}\@footnotemark}
\makeatother
\begin{document}
\date{}

\newcommand{\solution}{SocChain\xspace}
\newcommand{\solutionlong}{Social Choice Blockchain\xspace}
\newcommand{\Middleware}{Middleware\xspace}
\newcommand{\middleware}{middleware\xspace}
\newcommand{\consensus}{DBFT\xspace}

\title{\solution: Blockchain with Swift Proportional Governance for Bribery Mitigation}

\author{Deepal Tennakoon, Vincent Gramoli}
\email{dten6395@uni.sydney.edu.au, vincent.gramoli@sydney.edu.au}
\affiliation{%
\institution{University of Sydney}
\department{School of Computer Science}
\country{Australia}}

\begin{abstract}

Blockchain governance is paramount to lead securely a large group of users towards the same goal without disputes about the legitimacy of a blockchain instance over another.
As of today, there is no efficient way of protecting this governance against an oligarchy.
This paper aims to offer a new dimension to the security of blockchains by defining the \emph{swift proportional governance} problem. This problem is to rapidly elect governance users that proportionally represent voters without the risk of dictatorship. 
We then design and implement an open permissioned
blockchain called \solution (\solutionlong) that mitigates bribery by building upon results in social choice theory.
We deploy \solution and 
evaluate 
our new multi-winner election DApp running on top of it.
Our results indicate that, using our DApp, 150 voters can elect a proportionally representative committee of 150 members within 5 minutes. 
Hence we show that \solution can elect as many representatives as members in various global organizations. 

\end{abstract}

\ccsdesc[500]{Security and privacy~Distributed systems security}
\maketitle
\pagestyle{plain}

\vspace{-1em}
\section{Introduction}

\sloppy{This paper aims to offer a new dimension to the security of blockchain~\cite{Nak08} by introducing a framework
to define, solve and evaluate a \emph{swift proportional governance} as the problem of electing rapidly a committee that proportionally represents voters, to cope with bribery and dictatorship.

The notion of \emph{governance}, which is generally understood as the processes relied upon to make decisions 
and modify the protocol, has become an important topic in blockchain~\cite{Fli18,Zam19,BCC20}.  
The absence of governance already led users to create dissident instances of the two largest blockchains~\cite{KLM17,Web18}.
The worst thing that can happen 
is when an attacker takes control of the governance, which is best addressed through decentralization. 
Recent efforts were already devoted to applying social choice theory to distributed systems to cope with a fixed coalition of malicious users, also called a byzantine oligarchy~\cite{ZSC20}.
Yet, such solutions do not rotate the governance and fail as soon as the attacker manages to bribe a third of the governance.

A pernicious threat is thus the risk of obtaining an oligarchy that acts as a dictator.
The Proof-of-Stake (PoS) design that favors wealthy participants over others gained 
popularity as an efficient replacement to Proof-of-Work (PoW) in blockchain designs.
Combined with the Pareto Principle~\cite{Par64} stating that few users own most of the resources of the system or with bribery as the act of offering something to corrupt a participant, the system may end up being governed by an oligarchy.
Of course, no blockchains can be implemented if an adversary is capable of bribing all 
nodes instantaneously, this is why a slowly adaptive adversary is generally assumed by blockchains~\cite{GHM17,Rchain,8418625}.
Assuming that bribing takes time is reasonable~\cite{Lui85}:
Typically, a user can easily bribe a close friend but will take more time to bribe an acquaintance and may even fail at bribing a stranger (due to the fear of being exposed as corruptible). 
But even under this assumption,
the risk of an oligarchy remains.

Hence, blockchains require a fast governance reconfiguration that counteracts a growing coalition of malicious nodes by selecting a diverse 
and slow-to-bribe set of governance users, a problem we call \emph{swift proportional governance}.

The first part of this problem is to select a \uline{diverse set of governance users} 
or \emph{governors} that represent proportionally the voters to prevent an adversary, who controls $f<n/3$ of the $n$ governors, from acting as a dictator.
This ratio comes from (i)~the need for voters to reach consensus on the new set of governors and (ii)~the impossibility of solving consensus with $f\geq n/3$ malicious participants in the general setting~\cite{PSL80}.

One may think of reconfiguring the governance by executing the byzantine fault tolerant (BFT) consensus protocol of a blockchain not to decide upon a new block but to decide a set of governors~\cite{CCC20}. Most blockchain consensus protocols are, however, designed to offer a single-winner election: they are tuned to pick one block out of many legitimate blocks.
To make things worse, this picked block in consensus protocols is typically imposed by a winner/leader node~\cite{Nak08,Woo15,Kwo15,pass2017hybrid,BSK18,BBC19,ZILLIQA} that acts as a dictator.
For governance, we need instead a multi-winner election protocol so that voters can rank candidates, and the protocol outputs a set of candidates representative of the voted preferences. 
An example of a multi-winner election protocol is the Single Transferable Vote (STV) protocol~\cite{Tid95}, used for example to elect the Australian senate~\cite{ecanz21},
that can transfer each vote between candidates in the order of the voter's preferences.
However, this protocol is synchronous 
and costly~\cite{bartholdi1991single} to run within a consensus algorithm.
Another approach is thus to implement STV
in a smart contract: provided that the blockchain is consistent, the output of the smart contract should be the same across users, without the need for an additional consensus step.

\sloppy{The second part of the problem is a \uline{fast governance reconfiguration}: the longer a proposed governance update takes to be agreed upon, the greater the risk of the governance being bribed.
With an average latency of minutes~\cite{Woo15}
or an hour~\cite{Nak08} to commit a transaction agreed by all, blockchains are often subject to congestion when the demand rises~\cite{SFG19}. This congestion 
would also delay the execution of a smart contract intended to update the governance.
The recent performance improvements of open blockchains~\cite{GHM17,LLM19,Eth2,CNG21} relying on a subset of permissioned service providers to run consensus seems promising for reconfiguration~\cite{pass2017hybrid,AMN0S17,VG19,BAS20}.
One of the most recent of these blockchains even offers the finality of world-wide transactions within 3 seconds on average~\cite{CNG21} but does not support smart contract executions.
The absence of fast smart contract executions is not the only impediment.
Once the execution terminates, the blockchain service governed by the old configuration has to be shut down before the blockchain service with the new configuration can be started. Besides the downtime, the users who do not shut down their blockchain service would create a split, again leading users to create dissident instances of the same blockchain~\cite{KLM17,Web18}.
Instead, if all users initially joining agree that the blockchain self-reconfigures upon a special smart contract execution, then no split can occur. To achieve this, we need a new blockchain that reconfigures
its governors based on the smart contract output.}

\begin{table}
	\setlength\tabcolsep{1pt}
	\center
	\resizebox{\columnwidth}{!}{
	\begin{tabular}{r|lcc}
		\toprule
		Blockchain & Election & Proportionality & Non-dictatorship \\
		\midrule
		Tendermint~\cite{Kwo14} &  \cellcolor{tomato}None &  \cellcolor{tomato}no &  \cellcolor{tomato}no \\
		Algorand~\cite{GHM17} & \cellcolor{lightred}Sortition &  \cellcolor{tomato}no & \cellcolor{tomato}no\\
		Hybrid consensus~\cite{pass2017hybrid} & \cellcolor{lightred}PoW puzzle &  \cellcolor{tomato}no & \cellcolor{tomato}no\\
		Zilliqa~\cite{ZILLIQA} & \cellcolor{lightred}PoW puzzle &  \cellcolor{tomato}no & \cellcolor{tomato}no\\
		OmniLedger~\cite{8418625} & \cellcolor{lightred}Sortition & \cellcolor{tomato}no & \cellcolor{tomato}no\\
		RapidChain~\cite{Rchain} & \cellcolor{lightred}PoW puzzle & \cellcolor{tomato}no & \cellcolor{tomato}no\\
		ComChain~\cite{VG19} & \cellcolor{tomato}None &  \cellcolor{tomato}no &  \cellcolor{tomato}no \\
		Libra~\cite{BBC19} &  \cellcolor{tomato}None &  \cellcolor{tomato}no &  \cellcolor{tomato}no \\
		SmartChain~\cite{BAS20} &  \cellcolor{tomato}None &  \cellcolor{tomato}no &  \cellcolor{tomato}no \\
		Polkadot~\cite{cevallos2020verifiably} & \cellcolor{lightred}Multi-winner approval voting & \cellcolor{lightred}yes$^{*}$ & \cellcolor{tomato}no\\
		EOS~\cite{EOS-DPOS} & \cellcolor{lightred}Multi-winner approval voting & \cellcolor{lightred}yes$^{*}$ & \cellcolor{tomato}no\\
		SocChain & \cellcolor{lightlightgreen}Multi-winner preferential voting & \cellcolor{lightlightgreen}yes & \cellcolor{lightlightgreen}yes\\
		\bottomrule
	\end{tabular}}
	\begin{flushleft}
	* Polkadot and EOS offer some form of ``proportionality'' but do not satisfy the traditional definition we use~\cite{Woo94}.
	\end{flushleft}
	\caption{Blockchains with reconfigurable governance do not offer both proportionality and non-dictatorship while electing governors\label{table:comparison}
	}
	\vspace{-2.9em}
\end{table}

This paper defines the swift proportional governance problem (\cref{sec:pb}), designs a solution for it, proves the solution correct and evaluates the solution. 
Our proposed solution offers two practical contributions: 
an election decentralized application, or \emph{DApp},
that elects a set of governors ensuring proportionality and non-dictatorship (\cref{sec:stv}) and a blockchain (\cref{sec:solution}) that swiftly replaces a set of governors by this newly elected one. 
In particular, Table~\ref{table:comparison} indicates why other blockchain governance protocols do not address 
the same problem (the detailed comparison is deferred to \cref{sec:rw}).
The problem our election protocol solves 
NP-hard (\cref{ssec:complexity}) and, as we explain in \cref{sec:scalability}, it would be too slow to cope with bribery if executed on another blockchain. 
More specifically, our contributions are as follows:
\begin{itemize}
\item We introduce the first byzantine fault tolerant multi-winner election protocol, called \emph{BFT-STV}, 
a new primitive that augments the STV election procedure to enforce non-dictatorship in the presence of 
at most $t<n/3$ byzantine voters among $n$ voters without assuming synchrony (we denote by $f\leq t$ the actual number of byzantine voters).  
We implement this new protocol in a smart contract written in the Solidity programming language to allow the users of a blockchain to propose and rank 
candidates in the order of their preferences. 
As it is impossible to distinguish a non-responsive byzantine voter from a delayed message, we introduce
a new election quota $q_B = \frac{n-t}{k+1}$ where $k$ is the size of the committee.
Interestingly, we show that our new BFT-STV protocol preserves the proportionality and non-dictatorship properties of STV while ensuring termination.

\item This smart contract alone is not sufficient to ensure the swiftness of the governance reconfiguration, especially with Ethereum.
Our second contribution is a blockchain, called \solution (\solutionlong),  
that reconfigures itself by taking as an input the elected committee of governors output by the smart contract.
Similarly to the ``open permissioned''  Red Belly Blockchain~\cite{CNG21}, \solution accepts permissionless clients to issue transactions that permissioned governors agree upon.
The key difference is that \solution embeds the Ethereum Virtual Machine
supporting smart contracts that
can modify, at runtime, the set of permissioned nodes governing the protocol.
It then reconfigures fast the blockchain nodes in order to mitigate bribery attacks.
In particular, our protocol revokes permissions of existing governors to select new governors before a large portion of them could be bribed.
This is done by changing the governor set periodically and rapidly by electing new governors. 

\item We prove that our protocols are correct and evaluate the time they take to reconfigure a blockchain with up to 150 voters electing governors among 150
candidates. Our results indicate that it always takes less than 5 minutes for \solution to elect new governors and transition from 
using the old governors to using the new governors that will produce the upcoming blocks. Finally, we also evaluate \solution at a larger scale, showing that it performs thousands of transactions per second when deployed on 100 VMs, 
hence being able to replicate the governance maintained by major global organizations
such as OECD, EU, the CommonWealth, APAC, which all have under 100 members. 
\end{itemize}

In the remainder of the paper, we present the background and motivations (\cref{sec:background}), 
and our goal, model, and problem definition (\cref{sec:pb}).
We present our new secure governance DApp (\cref{sec:stv}) and prove it correct. 
We then present \solution (\cref{sec:solution}), and evaluate it with our new secure governance DApp
(\cref{sec:evaluation}). 
Finally, we present the related work (\cref{sec:rw}) and conclude (\cref{sec:conclusion}).
We defer the proof of correctness of our blockchain (\cref{line:proof}), the discussion of our solution (\cref{sec:discussion}) and the Solidity code of the BFT-STV smart contract (\cref{app:sc}) to the optional 
appendix.

\section{Background and Motivations}\label{sec:background}

The notion of governance, which is the processes relied on to make decisions impacting the protocol
has become 
an important topic in blockchain~\cite{Fli18,Zam19,BCC20}.
The governance structure encompasses the identity of parties capable of suggesting changes, the avenue through which such changes are proposed, the users capable of deciding the changes and the parties implementing these changes.
Due to the large number of users of a blockchain, governance is especially relevant to lead this large cohort towards a common goal.
With a lack of governance, the divergence of opinions may result in the split of the blockchain into multiple instances sharing a common transaction history but accepting distinct transactions.

\begin{figure}[t]
\begin{center}
\includegraphics[scale=0.35]{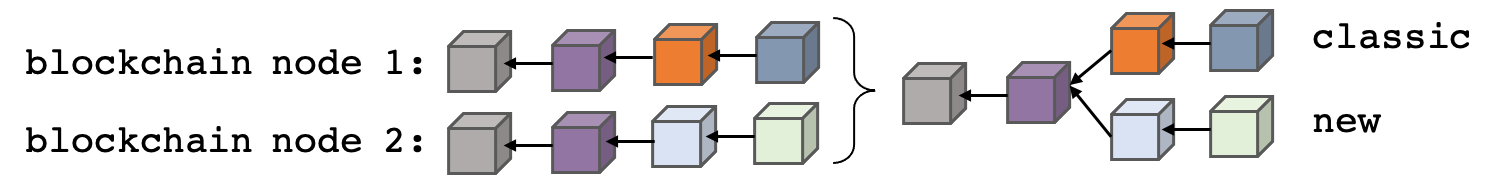}
\vspace{-1em}
\caption{If blockchain nodes disagree on a protocol update then they may start accepting distinct blocks, which results in a hard-fork with a classic version of the blockchain (e.g., ETC, BTC) and a new version of it (e.g., BCH, ETH). This can be avoided if the protocol includes at start-time a pre-determined procedure to reconfigure itself based on later governance decisions.\label{fig:fray}}
\end{center}
\vspace{-2em}
\end{figure}

As an example, consider Figure~\ref{fig:fray}, where blockchain node 1 rejects a software upgrade and keeps accepting old-formatted blocks whereas blockchain node 2 accepts this upgrade and starts accepting blocks in a new format, leading to a hard fork.
The two largest blockchains were victims of such splits: Bitcoin is now split into BTC and BCH~\cite{Web18} whereas Ethereum is now split into ETH and ETC~\cite{KLM17}.
The absence of governance can draw blockchain users into such clashes. 
The solution to this problem, which we adopt here, is to ``hard-code'' in the blockchain software a reconfiguration (or upgrade) that executes as soon as it is voted upon. 
When the blockchain is spawned for the first time, all its users implicitly accept that it may reconfigure. Later, if a majority of voters decide to reconfigure,
then the blockchain changes the software and its governance users, or \emph{governors}, automically.
There is no need for the users to decide whether to upgrade as their blockchain node reconfigures autonomously.

The biggest challenge is to prevent an attacker from obtaining the control of the governance,
which is usually tackled through decentralization. 
Recently, the best paper at OSDI 2020~\cite{ZSC20} 
proposes to apply social choice theory results to distributed systems in order to guarantee that no coalition of $f<n/3$ 
governors can dictate the order of transactions.
Its authors 
assume that the governors running the consensus are pre-determined and do not aim at running an election that will update this set of governors. 
%
The reason for this assumption stems from the conjunction of two fundamental results of distributed computing indicating that one cannot implement a secure blockchain 
as soon as the oligarchy includes $n/3$ participants because (i)~a blockchain needs to solve consensus in the general setting~\cite{GKM19} and (ii)~consensus cannot be solved if network delays cannot be predicted~\cite{FLP85}. 
As soon as more than $n/3$ of the governors fail, then the governance cannot lead participants towards the same goal anymore.

As blockchains typically handle valuable assets, several works already noted the risk for a user to 
bribe other users to build an oligarchy 
capable of stealing these assets~\cite{Bon16}.
As mentioned before, it is reasonable to assume that bribing many nodes is not instantaneous~\cite{GHM17,Rchain,8418625,Lui85}.
To reduce the chances that governance users, or \emph{governors}, know each other,
Algorand~\cite{GHM17} exploits randomness and non-interactiveness~\cite{algorand21},
however, a random selection does not eradicate the possibility of obtaining a byzantine oligarchy, because in Algorand the more coins users have, the higher their chances of being selected.
Other blockchains assume exlicitly a \emph{slowly-adaptive} adversary~\cite{LNZ16,8418625}, 
assuming that the adversary can corrupt a limited number of nodes on the fly at
the beginning of each consensus epoch but cannot change the set of malicious participants during an epoch.
We build upon such an assumption to implement \solution (\cref{sec:solution}).

Traditional blockchains do not offer a representative governance~\cite{Bla58}. 
Some blockchains give permissions to miners to decide to change the gas price~\cite{SFG19}, others give the permission of deciding a new block randomly~\cite{GHM17}, some prevent governors from changing the 
governance size~\cite{EOS}.
The closest work, concomitant with ours and part of Pokaldot~\cite{BCC20}, targets proportional representation while favoring the wealthiest users by offering an approval voting system.
Given the Pareto Principle~\cite{Par64} stating that few users typically own most of the resources, care is needed to avoid falling back to an oligarchy.
In order to pursue the
two conflicting goals of letting the wealthiest participants govern while trying to avoid that they constitute an oligarchy, Polkadot can only offer
some approximation to the problem solution~\cite{CS20b}. 
Instead of approximating a solution that could result in the blockchain being unusable, 
we offer a preferential voting system that solves exactly the problem of proportional representation and non-dictatorship as we explain in~\cref{ssec:gov}.

\subsection{Social choice theory with byzantine fault tolerance}\label{sec:socialchoice}


To propose meaningful properties for blockchain governance, we draw inspiration from classic work on social choice theory.
Given a set of $n$ voters, each casting an \emph{ordinal ballot} as a preference order over all $m$ candidates, a \emph{multi-winner election} protocol outputs a winning committee of size $k$.

Arrow~\cite{Arr50} defined \emph{non-dictatorship} as a property of a voting protocol where there is no single person that can impose its preferences on all. Our goal is to adapt this property to cope with byzantine voters such that non-dictatorship remains satisfied even when an adversarial person controls up to $f<n/3$ byzantine voters (Def~\ref{def:governance}).

Non-dictatorship is however insufficient to guarantee that newly elected governors remains a diverse 
representation of the voters.
Black~\cite{Bla58} was the first to define this proportionality problem where elected members must represent ``all shades of political opinion'' of a society.

Dummett~\cite{Dum84} introduced \emph{fully proportional representation} to account for ordinal ballots, containing multiple preferences.
Given a set of $n$ voters aiming at electing a committee of $k$ governors, if there exists $0<\ell \leq k$ and a group of $\ell\cdot q_H$ who all rank the same $\ell$ candidates on top of their preference orders, then these $\ell$ candidates should all be elected. 
However, it builds upon Hare's quota $q_H$, which is vulnerable to strategic voting whereby a majority of voters can elect a minority of seats~\cite{LH07}.
This problem was solved with the introduction of Droop's quota $q_D$ as the smallest quota such that no more candidates can  be  elected  than  there  are  seats  to fill~\cite{Tid95}.
%


Woodall~\cite{Woo94} replaces Hare's quota with Droop's quota $q=\lfloor \frac{n}{k+1} \rfloor$ and defines the \emph{Droop proportionality criterion} as a variant of the fully proportional representation property:
if for some whole numbers $j$ and $s$ satisfying $0 < j \leq s$, more than $j \cdot q_D$ of voters put the same $s$ candidates (not necessarily in the same order) as the top candidates in their preference list, then at least $j$ of those $s$ candidates should be elected.
This is the property we target in this paper and we simply rename it  \emph{proportionality}
(Def.\ref{def:governance}).


It is known that the First-Past-The-Post (FPTP) single-winner election and the Single Non-Transferrable Vote (SNTV) multi-winner election cannot ensure fully proportional representation~\cite{FSS17}. The reason is that 
voters can only reveal their highest preference. 

This property can however be achieved using the \emph{Single Transferable Vote (STV)} algorithm with Hare's quota $q_H = \frac{n}{k}$. 
In STV, candidates are added one by one to the winning committee and removed from the ballots if they obtain a quota $q$ of votes. 
STV is used to elect the Australian senate and is known to 
ensure fully proportional representation.
Unfortunately, this protocol is synchronous~\cite{DLS88} 
in that its quotas generally rely on the number of votes $n$ received within a maximum voting period.

As one cannot predict the time it will take to deliver any message on the Internet, 
one cannot distinguish a slow voter from a byzantine one.
Considering $n$ as the number of governors or potential voters among which up to $t$ can be bribed
or byzantine, our protocol can only wait for at most $n-t$ votes to progress without assuming synchrony.
Waiting for $n-t$ prevents us from guaranteeing that the aforementioned quotas can be reached.
We thus
define a new quota called the \emph{byzantine quota} $q_B=\lfloor \frac{n-t}{k+1} \rfloor$ such that $t<n/3$ and reduce the number of needed votes to start the election to $n-t$. 
Of course, up to $t$ of these $n-t$ ballots may be cast by byzantine nodes, 
however, we show in Theorem~\ref{thm:non-dictatorship} that no adversary controlling up to $t$ byzantine nodes can act as a dictator.
Based on $q_B$, we propose BFT-STV that extends STV for a byzantine fault tolerance environment. We also show that BFT-STV satisfies proportionality and non-dictatorship (\cref{sec:stv}) without assuming synchrony.

\section{The Swift Proportional Governance Problem}\label{sec:pb}

Our goal is to offer swift proportional governance
by: (i)~offering a blockchain governance that allows distributed users to elect a committee proportionally representative of the voters and without dictatorship and 
(ii)~guaranteeing security of the blockchain by changing rapidly its governance. 
%
%
We first present 
the computation model (\cref{ssec:model}) before defining the BFT governance (\cref{ssec:gov}) and blockchain (\cref{ssec:bc}) 
problems separately, and terminate with the threat model (\cref{ssec:threat}).

\subsection{Byzantine fault tolerant distributed model}\label{ssec:model}

We consider a distributed system of $n$ nodes, identified by public keys $I$ and network identifiers (e.g., domain names or static IP addresses) $A$,
that can run different services: (i)~the state service
executes the transactions and maintains a local copy of the state of the blockchain, (ii)~the consensus service 
executes the consensus protocol in order to agree on a unique block to be appended to the chain.
Client\footnote{The term ``client'' is often used in Ethereum to refer to a node regardless of whether it acts as a server. We use client in the traditional sense of the client-server distinction~\cite{TvS07}.} 
nodes simply send transaction requests to read from the blockchain (to check an account balance) or 
to transfer assets, upload a smart contract or invoke a 
smart contract\footnote{Note that some smart contract invocations are considered read-only, we do not distinguish them from those updating for simplicity in the presentation.}.

As we target a secure blockchain system running over an open network like the Internet, we
consider the strongest fault model called the byzantine model~\cite{LSP82}, where nodes can fail arbitrarily by, for example, sending erroneous messages
and we do not assume that the time it takes to deliver a message is upper bounded by a known delay, instead 
we assume that this delay is unknown, a property called \emph{partial synchrony}~\cite{DLS88}.
We also aim at implementing an optimally resilient system: as blockchain requires consensus in the general model~\cite{GKM19} and consensus cannot be solved in the partially synchronous model with $n/3$ byzantine nodes~\cite{LSP82}, we assume a slowly adaptive byzantine adversary where the number $f$ of byzantine 
governors can grow up to $t<n/3$ within the first $\Delta$ units of time of the committee existence 
(we will show in \cref{ssec:setup} how $\Delta$ can be made as low as 5 minutes). A node that is not byzantine is called \emph{correct}. 
%
Finally, we assume public key cryptography and that the adversary is computationally bounded.
Hence, the issuer of a transaction can \emph{sign} it and any recipient can correctly verify the signature.

\subsection{Secure governance problem}\label{ssec:gov}
We refer to the blockchain governance problem as the problem
of designing a BFT voting protocol in which $n$ voters rank $m$ candidates to elect a committee of $k$ governors ($k < m \leq n$) to ensure non-dictatorship as defined by Arrow~\cite{Arr50} and proportionality as defined by Dummett~\cite{Dum84}, Woodland~\cite{Woo94} and Elkind et al.~\cite{EFS17} (cf. \cref{sec:socialchoice}). The main distinction is that we adapt this problem from social choice theory to the context of distributed computing.
\begin{definition}[The Secure Governance Problem]\label{def:governance}
The \emph{secure governance problem} is for a distributed set of $n$ voters, among which $f\leq t <n/3$ are byzantine, to elect a winning committee of $k$ governors among $m$ candidates (i.e., $m>k$) such that the two following properties hold:
\begin{itemize}
\item \label{prop-criteria} \emph{Proportionality:} if, for some whole numbers $j$, $s$, and $k$ satisfying $0 < j \leq s \leq k$, more than $j(n-t)/(k+1)$ of voters put the same $s$ candidates (not necessarily in the same order) as the top $s$ candidates in their preference listings, then at least $j$ of those $s$ candidates should be elected.
\item \label{non-dictatorship} \emph{Non-dictatorship:} a single adversary, controlling up to $f<n/3$ byzantine voters, cannot always impose their individual preference as the election outcome. 
\end{itemize}
\end{definition}
The need for these two properties stems from our goal of guaranteeing 
proportional representation (proportionality) but also disallowing a coalition of byzantine nodes from imposing their decision on the rest of the system (non-dictatorship).
Note that the non-dictatorship property differs slightly from the original definition~\cite{Arr50} that did not consider a byzantine coalition. 
In particular, our property considers coalitions and prevents them from imposing their preference in ``all'' cases.

\subsection{Blockchain problem}\label{ssec:bc}

We refer to the blockchain problem as the problem of ensuring both the
safety and liveness properties that were defined in the literature by Garay et al.~\cite{GKL15} and restated more recently by Chan et al.~\cite{CS20}, and a classic validity property~\cite{CNG21} to avoid trivial solutions to this problem.
\begin{definition}[The Blockchain Problem]\label{def:blockchain}
The \emph{blockchain problem} is to ensure that a distributed set of blockchain nodes 
maintain a sequence of transaction blocks such that the following properties hold:
\begin{itemize}
\item  \emph{Liveness:} if a correct blockchain node receives a transaction, then this transaction will eventually be reliably stored in the block sequence of all correct blockchain nodes.
\item \emph{Safety:} the two chains of blocks maintained locally by two correct blockchain nodes are either identical or one is a prefix of the other. 

\item \emph{Validity:} each block appended to the blockchain of each correct blockchain node is a set of \emph{valid} transactions (non-conflicting well-formed transactions that are correctly signed by their issuer).
\end{itemize}
\end{definition}
The safety property does not require correct blockchain nodes to share the same copy, simply because one replica may already have received the latest block before another receives it.
Note that, as in classic definitions~\cite{GKL15,CS20}, the liveness property does not guarantee that a client transaction is included in the blockchain: if a client sends its transaction request exclusively to byzantine nodes then byzantine nodes may decide to ignore it.


\remove{
\subsection{System Model}\label{ssec:sysmodel}
In Figure~\ref{fig:UseCase}, we present the system model in brief. Our system model consists of a voting DApp deployed and two main actors\vincent{I would not introduce the term actors, we have too many terms already, let's simplify}. This model is later deployed in \solution.
\begin{itemize}
	\item Voters/Governors - Casts ordinal ballots on candidates to a blockchain smart contract. Additionally, they proposers blocks and takes part in consensus of \solution.
	\item Candidates - Are included in the ballot preferences and have the goal of being elected as the next set of governors
	\item Governors (committee members) - Proposers blocks and takes part in consensus \vincent{No, we agreed that voters=governors.}\deepal{Yes, we can make voters=governors, I have changed the first actor accordingly, but in our experiment Fig. 6, we elect governors=candidates/2, so voters=governors was not always the case. Can we say we assume voters=governors here but for the experiment we assumed it was not the case?}
\end{itemize}

\begin{figure}[t]
\begin{center}
\includegraphics[scale=0.38]{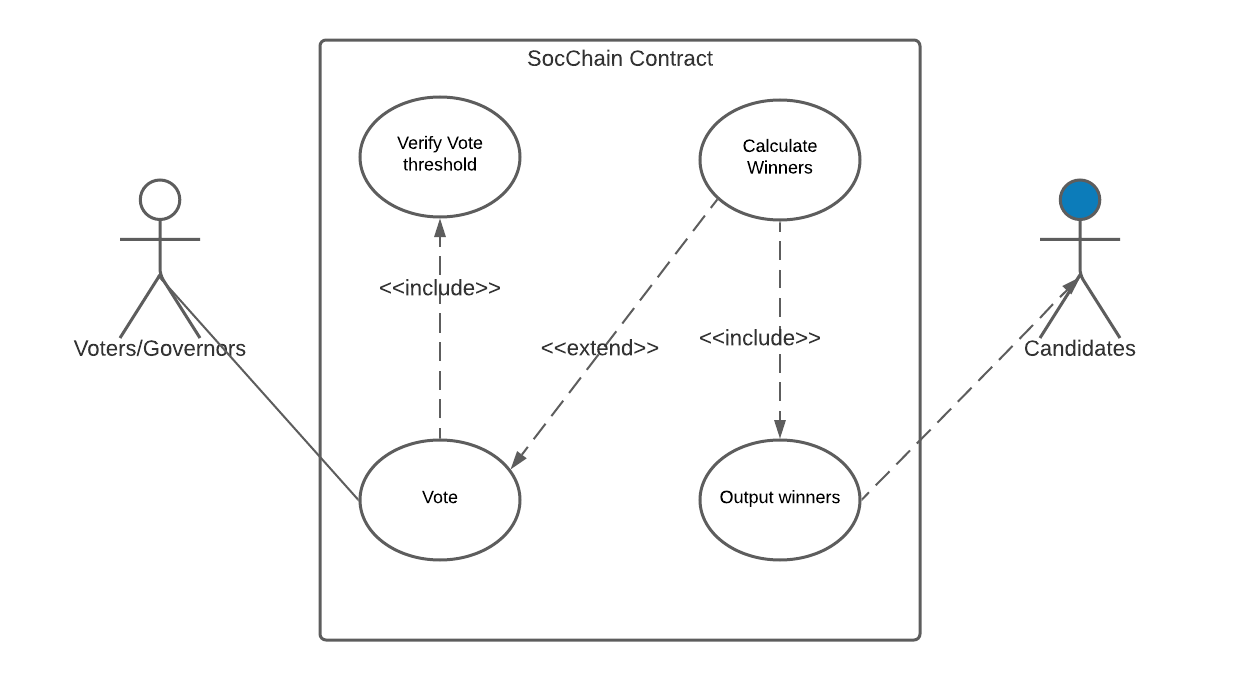}
\caption{The use case diagram for our system.}
\end{center}
\end{figure}
}
\subsection{Threat model}\label{ssec:threat}
\remove{\deepal{In this section, we present our threat model. Note that it is hard or almost impossible to mitigate every threat in a system without impacting usability. There is always a trade-off and hence we only focus on a few obvious and well-known threats to BFT blockchains like our \solution.

\begin{table}[H]
	\setlength\tabcolsep{1pt}
	\center
	\resizebox{\columnwidth}{!}{
		\begin{tabular}{c|c|c|c|c}
			\toprule
			Threat & Threat type & Asset & Threat Actor &  Mitigation  \\
			\midrule
			Bribery attack & Tampering \solution & Governor &  Swift Proportional Secure Governance \\
			\midrule
			Sybil attack & Spoofing & \solution & Blockchain node & KYC (Know Your Customer) \\
			\bottomrule
	\end{tabular}}
	\caption{The threat model\vincent{this table is unclear and incomplete, just remove it}\label{table:threatmodel}
	}
\end{table}
}
}
As in previous blockchain work~\cite{LNZ16,GHM17,Rchain,8418625}, 
we assume a slowly adaptive adversary with a limited bribing power that cannot, for example, bribe all users instantaneously.  More precisely, 
provided that any new set of governors is elected with proportional representation, we also assume that it takes more than $\Delta = 5$ minutes 
for $1/3$ of new governors to misbehave as part of the same coalition. 
We will show in \cref{ssec:setup} that $\Delta = 5$ minutes is sufficient once the votes are cast as \solution reconfigures its governance in less than 5 minutes. 
In comparison, once it will be available, Eth2.0 will take at least 6.4 minutes to reconfigure governance~\cite{wels2019guaranteed}.

For the initial set of governors to be sufficiently diverse, we can simply select governors based on their 
detailed information. This can be done by requesting initial candidates to go through a 
\emph{Know-Your-Customer (KYC)} identification process, similar to the personal information requested by the Ethereum proof-of-authority network to physical users before they can run a validator node~\cite{poa}.
A set of governors could then be selected depending on the provided information by making sure multiple governors are not from the same jurisdiction, they are not employed by the same company, they represent various ethnicities, they are of balanced genders, etc.
We defer the details of how the KYC process can be implemented, how user anonymity can be preserved and 
how to cope with bribery smart contract attacks in \cref{app:anonymity}.

\subsubsection{Bribery attack}
Limiting the number of nodes responsible to offer the blockchain service as done in recent open blockchains~\cite{CNG21}
exposes the service to a
bribery attack~\cite{Bon16}, which is an act of offering something to corrupt a participant.
This is because it is typically easier to bribe fewer participants.
In particular, 
as consensus cannot be solved with at least $\frac{n}{3}$ byzantine processes among $n$ when message delays
are unknown~\cite{DLS88}, it is sufficient to bribe $\frac{n}{3}$ processes 
to lead correct blockchain nodes to disagree on the next block appended to the blockchain and thus create a fork in the blockchain.
The attacker can then exploit this fork to have its transaction discarded by the system and then re-spend the assets he supposedly transferred 
in what is called a \emph{double spending}. 
Our reconfiguration protocol mitigates such a bribery attack in the presence of a slowly-adaptive adversary by re-electing $n$ new governors that execute the consensus protocol
every $x$ blocks (by default we use $x=100$). This is how we prevent the risks that $\frac{n}{3}$ of the current governors get bribed when the blockchain has between $k$ and $k+x$ blocks.
As $x$ consecutive block creations do not always translate into the same time interval, 
we detail in~\cref{sec:discussion} how one can make sure that, periodically, exactly $x$ blocks are created.

\subsubsection{Sybil attacks}\label{sec:sybil}
A Sybil attack consists of impersonating multiple identities to overwhelm the system---in the context of votes, a Sybil attack could result in double voting. 
The traditional blockchain solution, proof-of-work~\cite{Nak08}, copes with Sybil attacks by requiring each block to include the proof of a crypto puzzle.
Proof-of-stake, give permissions to propose blocks to the wealthiest participants by relying on the assumption that participants with a large stake in the system behave correctly.
We adopt a third solution that consists of providing 
authenticating information, in the form of know-your-customer (KYC) data, in exchange for the permission to propose new blocks, vote for governors, or be a governor candidate.
This authentication copes with Sybil attacks by preventing the same authenticated user from using distinct node identities (as detailed in~\cref{sec:discussion}). 


\section{Byzantine Fault Tolerant Proportional Governance}\label{sec:stv}

In this section, we present how 
to elect, despite $f\leq t<n/3$ byzantine nodes, a diverse set of governors. The idea is to allow a set of $n$ blockchain nodes that vote to elect a committee proportionally representing the voters.
To this end, we propose the \emph{Byzantine Fault Tolerant Single Transferrable Vote (BFT-STV)} smart contract that implements a multi-winner election that solves the governance problem (Def.~\ref{def:governance}).
We detail how to integrate it into \solution in~\cref{sec:solution}. 

\begin{figure}[t]
\begin{center}
\includegraphics[scale=0.38]{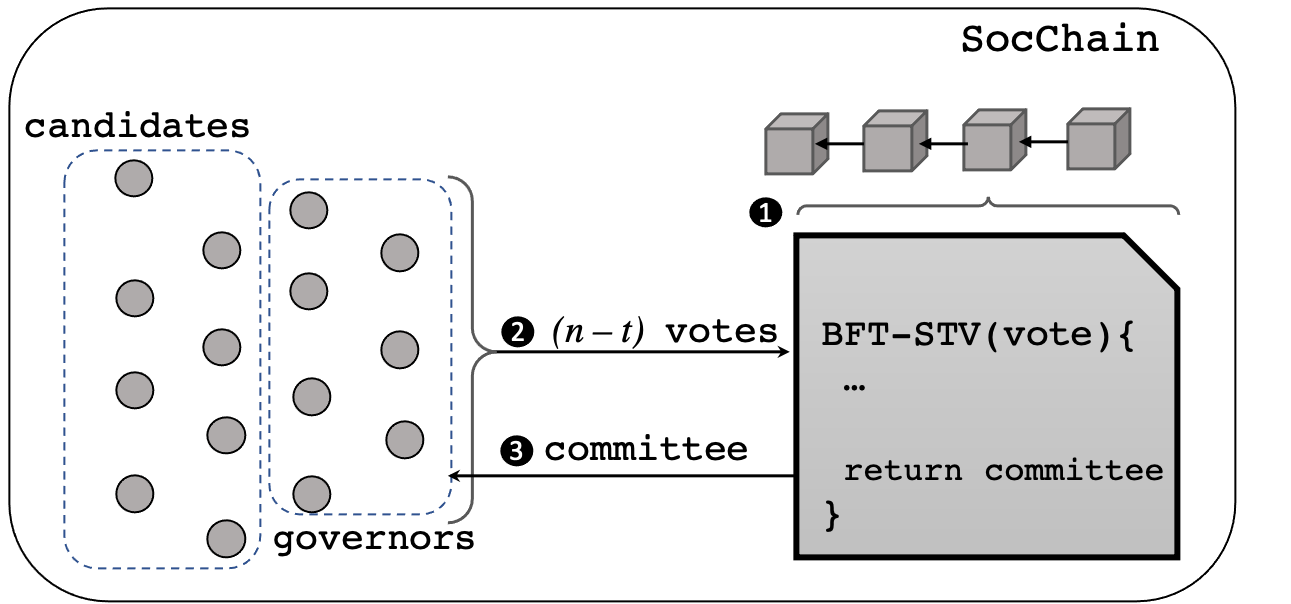}
\caption{
The smart contract that implements the BFT-STV protocol is on-chain \ding{202}, takes as an input a set of at least $(n-t)$ ballots (each ranking $k$ candidates among $m$) cast by $(n-t)$ voters among the $n$ governors \ding{203} 
and outputs a committee of $k$ elected nodes \ding{204} to play the role of the new governors. 
Note  
that the last committee of nodes elected to play the role of governors will then vote for the next committee \ding{203} and so on (one can fix $k=n$ so that the committee size never changes).
\label{fig:bftstv} 
}
\end{center}
\end{figure}

\subsection{Overview}

In order to guarantee that the election ensures fully proportional representation, we designed the BFT-STV algorithm and implemented it
in a smart contract. In this section, we present its high level pseudocode and defer the details of its implementation to \cref{app:sc}.
To bootstrap, the initial permissions to vote are obtained by $n$ initial governors after identification (KYC) to ensure diversity and prevent Sybil attacks (\cref{sec:sybil}).
Recall that governors cannot use the classic STV algorithm to elect a new committee as the smart contract has to progress despite up to $t < n/3$ byzantine voters not casting proper ballots and as the upper-bound on the message delay is unpredictable.
As depicted in Figure~\ref{fig:bftstv}, the BFT-STV smart contract takes, instead, as an input $n-t$ ballots cast by the voters. 
Each ballot consists of a rank of all the candidates, hence the name \emph{ordinal ballot}.
Once the threshold $n-t$ of cast ballots is reached, the BFT-STV contract selects the governors based on the preference order indicated in the $n-t$ ballots.
Traditionally, the STV algorithm consists of counting which candidates received a number of votes that exceed the quota $q_D = \frac{n}{k+1}$ where $k$ is the size of the  committee to be elected. However, as there can be at most $t$ byzantine nodes among the voters, we introduce the byzantine quota $q_B = \frac{n-t}{k+1}$ (denoted $q$ when clear from the context).
We will explain how the blockchain replaces the current governors by this newly elected committee of governors in~\cref{sec:solution}.

\begin{algorithm}[h!]
	\caption{Byzantine Fault Tolerant Single Transferable Vote (BFT-STV) - Part 1\label{alg:stv}}
	{\footnotesize
	\begin{algorithmic}[1]
        \Part{Initial state}
          \State $k \in {\mathbb N}$, the size of the targeted committee.
          \State $n\in {\mathbb N}$, the number of voters.
          \State $t\in {\mathbb N}$, an upper bound on the number $f$ of byzantine replicas, $f\geq t$.
          \State $m$, the number of candidates per ballot.
          \State $v$, a mapping from candidates to their number of votes.
          \State $\ms{ballots}$, the set of received ordinal ballots, initially $\emptyset$
          \State $C \subseteq I$, the set of candidates.
	 \State $\ms{E} \subseteq C$ the set of eliminated candidates, initially $\emptyset$.
	 \State $\ms{S} \subseteq C$ the set of winning candidates, initially $\emptyset$.
	 \State $pref[ballot]=index$ a mapping of ballot and its current preference index

        \EndPart
        
        \Statex
        
        \Part{$\lit{cast-ballot}(b)$} \Comment{cast ballot $b$}\label{line:castballot}
          \If{$\ms{well-formed}(b)$} $\ms{ballots} \gets \ms{ballots} \cup \{\ms{b}\}$ \Comment{store ballot}\label{line:wellformed}
          \EndIf
          \If{($\ms{ballots}$ has $n-t$ ballots from distinct voters)}   \label{line:nmtballots}\Comment{enough ballots} 
        	    \State $\lit{change-committee}(\ms{ballots})$\label{line:changecommittee}  \Comment{replace committee} 
        \EndIf
        
        \EndPart

        \Statex 
        
        \Part{$\lit{change-committee}(\ms{ballots})$} \Comment{replace committee}
          \ForAll{$b \in \ms{ballots}$}  \Comment{for each received ballot}\label{start-vote-count}
            \If{($\ms{b}[0]=c$ such that $c \in C$)}
               $\ms{v}[c] \gets \ms{v}[c] + 1$\Comment{\# 1st pref = $c$}\label{init-vote-count}
        	\EndIf
        	\State $\ms{pref}[b] \gets 0$ \Comment{assign pref. index of $b$ to the first preference/index 0}\label{pref-pointer}
        	\EndFor
        \State $\ms{round} \gets 0$ \Comment{first round}\label{round-start}
        \While{$(|S| < k)$}\Comment{until the new committee is full}\label{while-stv}
          \State$S \gets \lit{STV_{B}}(v, \ms{ballots}, pref)$ \Comment{invoke classic STV}\label{line:invokestv}
        	  \State $\ms{round} \gets \ms{round} +1$\label{line:round} \Comment{increment round number}\label{Increment-round}
        	  \If{$(|C|-|E| = k)$} $\lit{break}$ \Comment{stop eliminating}\label{line:stopelim}
        	  \EndIf
        \EndWhile
        
        \ForAll{$b \in \ms{ballots}$} \Comment{for each ballot}\label{extra-elect}
         \For{$(j = 0; j<m; j\text{++})$} \Comment{each candidate in decreasing pref. order}
          \If{$(|\ms{S}| < k \wedge b[j] \in C \setminus S \setminus E)$} \Comment{if eligible}
            \State $\ms{S} \gets \ms{S} \cup \{c\}$ \Comment{select $c$}\label{line:elect}
          \EndIf
         \EndFor
        \EndFor
        \State {\bf emit} $S$ \Comment{explicitly emit committee} \label{line:emit}
        
        
        \EndPart
        
        \algstore{stv}
        
	\end{algorithmic}%
}%
\end{algorithm}%
\begin{algorithm}[h!]
	\caption{Byzantine Fault Tolerant Single Transferable Vote  (BFT-STV) - Part 2\label{alg:stv1}}
	{\footnotesize
	\begin{algorithmic}[1]
	
        \algrestore{stv}

          \Part{Initial state}
          \State $k \in {\mathbb N}$, the size of the targeted committee.
          \State $n\in {\mathbb N}$, the number of voters.
          \State $t\in {\mathbb N}$, an upper bound on the number $f$ of byzantine replicas, $f\geq t$.
 	  \State $q_{B} = \frac{n-t}{k+1}$, the quota of votes to elect a candidate.
          \State $C \subseteq I$, the set of candidates.
	 \State $\ms{E} \subseteq C$, the set of eliminated candidates, initially $\emptyset$.
	 \State $\ms{S} \subseteq C$, the set of winning candidates, initially $\emptyset$.
	 \State $\ms{X} \subseteq C$, the set of excess candidates, initially $\emptyset$.
        \EndPart
        
        \Statex
        
        \Part{$\lit{STV}_B(v, \ms{ballots}, \ms{pref})$}\label{line:start-stv}
        \If{$\exists c \mid v[c] > q_{B}$}\Comment{if the quota is exceeded}\label{line:quota-exceed}
         \State $S \gets S \cup \{c\}$\Comment{elect candidate}
        
        \State $\ms{X}\gets \ms{X} \cup \{c\}$\Comment{save candidates that exceed quota in $\ms{X}$}
         \State$\ms{x[c]} \gets v[c] - q_{B}$\Comment{excess vote from candidate $c$}
        
        \ForAll{$b \in \ms{ballots}$} \Comment{for each ballot}        
        \If{$b[\ms{pref[b]}]=c$ and $c \in \ms{X}$}\Comment{if current ballot pref = one of X}
        \State$\ms{count[c]} \gets \ms{count[c]} + 1$\Comment{the number of candidates $c$}
        \State $\ms{pref-next[b]} \gets \ms{pref[b]} + 1$ \Comment{point to next preferred candidate}
        \While{$b[\ms{pref-next[b]}] \in (S \vee E)$}
        \Comment{while not uneligible}
        \State $\ms{pref-next[b]} \gets \ms{pref-next[b]} + 1$  \Comment{try next pref. pointer}
        \EndWhile
        
        \If{$b[\ms{pref-next[b]}] \not \in (S \cup E)$} \Comment{if eligible candidate found}
        \State $\ms{pref[b]} =\ms{pref-next[b]}$\Comment{move the preference pointer}
        \State $z \gets b[\ms{pref-next[b]}]$ \Comment{next preferred candidate in ballot}
        \State $\ms{cand-next} \gets \ms{cand-next} \cup \{\tup{c,z}\}$\Comment{current\&next candidates} 
        \State$\ms{count[z]} \gets \ms{count[z]} + 1$\Comment{The number of candidates $z$}
        \EndIf
        \EndIf
        \EndFor
        
        \ForAll{unique $\tup{c,z} \in \ms{cand-next}$} \Comment{transfer excess votes}
        \State$\ms{v[z]} \gets \ms{v[z]} + \ms{x[c]} \cdot (\ms{count}[z]/\ms{count[c]})$ \Comment{to next candidates}
\label{line:tranfer-excess} \label{line:stop-electing}
        \EndFor
        
        \EndIf
        
        \If{$\forall c : v[c] \le q_{B}$}\Comment{if no candidates exceed the quota in the round}\label{line:start-eliminating}
        \State $E \gets (E \cup t \mid t = \lit{min}_{\forall c}(v[c]))$\Comment{eliminate candidate with least votes}\label{line:eliminate}
        \State $\ms{transfer-vote} \gets v[t]$
        \State $v[t] \gets 0$\Comment{reset votes of least candidate to 0}

        \ForAll{$b \in \ms{ballots}$}
       	 	\While{$s<\ms{size}$}
                		\If{$b[s] = t$} \Comment{store  ballot and preference index...}
                			\State $\ms{elimpointer} \gets \ms{elimpointer}\cup (b,s)$ \Comment{...of least voted cand.}
                		\EndIf 
         		\State $s \gets s+1$ \Comment{Increment preference}
         	\EndWhile
         \EndFor
         
         \ForAll{$(b,s) \in \ms{elimpointer}$}
         \If{$b[s]=m \wedge m \in E$}\Comment{If preference $s$ of ballot $b$ is eliminated}
         \State $\ms{pref-next[b]} \gets s+1$
         \State $\ms{count[m]} \gets \ms{count[m]} + 1$\Comment{count of candidates $m$ in all ballots}
         \While{$b[\ms{pref-next[b]}] \in (S \vee E)$}\Comment{until  candidate is found}
         \State $\ms{pref-next[b]} \gets \ms{pref-next[b]} + 1$ \Comment{...increment pref. pointer}
         \EndWhile
         
         \If{$b[\ms{pref-next[b]}] \not\in  {S \cup E}$}
         \State $pref[b] \gets \ms{pref-next[b]}$\Comment{move the preference pointer}
         \State $z \gets b[\ms{pref-next[b]}]$ 
         \State $\ms{cand-next} \gets \ms{cand-next} \cup (m,z)$\Comment{least voted \& next cand.} 
         \State$\ms{count[z]} \gets \ms{count[z]} + 1$\Comment{the number of candidates $z$}
         \EndIf
         \EndIf
         \EndFor

        \ForAll{unique $(m,z) \in \ms{cand-next}$} \Comment{transfer from least voted cand.}
        \State$\ms{v[z]} \gets \ms{v[z]} + \ms{transfer-vote} \cdot (\ms{count[z]}/\ms{count[m]})$\label{line:transferelimnate}
        \EndFor
        
        \EndIf
        \State $X \gets null$ \label{line:stop-stv}
        \State {\bf return} $S$ \Comment{return the set of winning candidates} \label{line:emit2}
        \EndPart
	
	\end{algorithmic}%
}%
\end{algorithm}%

\subsection{Byzantine Fault Tolerant Single Transferrable Vote}\label{ssec:bftstv}

Algorithm~\ref{alg:stv} presents the main functions of the BFT-STV smart contract that the governors can invoke whereas Algorithm~\ref{alg:stv1} is the classic STV algorithm adapted to progress in a partially 
synchronous~\cite{DLS88} environment and despite the 
presence of up to $t$ byzantine voters, hence its name $\lit{STV}_B$.

Initially, the governors cast their ballots by invoking the function $\lit{cast-ballot(\cdot)}$ at line~\ref{line:castballot} of Algorithm~\ref{alg:stv}. As a result, the smart contract verifies that the ballots are well-formed (line~\ref{line:wellformed}). This involves checking that the governors have not voted for themselves on their ballots and there are no duplicated preferences. Although the details are deferred to the Appendix~\cref{app:sc} for simplicity in the presentation, note that the smart contract keeps track of the public keys of the governors casting ballots to ensure that the same governor cannot double vote. 
Once the smart contract receives $n-t$ well-formed ballots the $\lit{change-committee}(\cdot)$ function is invoked (line~\ref{line:changecommittee}).
The $\lit{change-committee}$ function starts by computing the score of the valid candidates as the number of votes they receive at lines~\ref{start-vote-count}--\ref{init-vote-count}. Valid candidates are initially selected through KYC (\cref{sec:sybil}) before being periodically voted upon by governors.
A preference pointer is initialized to the first preference of each ballot at  
line~\ref{pref-pointer}.
Then a new round of the STV election process starts (lines~\ref{round-start}--\ref{Increment-round}).
This execution stops once the committee of new governors is elected  (line~\ref{while-stv}).
If before the targeted committee is elected, the number of eliminated candidates has reached a maximum and no more candidates can be eliminated to achieve the target committee size, then the STV election stops (line~\ref{line:stopelim}). 
The remaining non-eliminated candidates are elected by decreasing order of preferences at lines~\ref{extra-elect}--\ref{line:elect} until the target committee size is reached.
Finally, the smart contract emits the committee of elected candidates (line~\ref{line:emit}), which notifies the replicas of the election outcome.

\subsection{Classic STV with the byzantine quota}

Algorithm~\ref{alg:stv1} presents the classic STV algorithm but using the new byzantine quota $q_B$ by electing candidates whose number of votes exceed $q_B$ (line~\ref{line:quota-exceed}). 
This algorithm executes two subsequent phases: in the first phase (lines~\ref{line:start-stv}--\ref{line:stop-electing}) the algorithm elects the candidates whose number of votes exceeds the quota $q_B = \frac{n-t}{k+1}$; in the second phase (lines~\ref{line:start-eliminating}--\ref{line:stop-stv}), the algorithm eliminates the least preferred candidate if no candidates received a number of votes that exceeds the quota.
%
In each round of STV function call (line~\ref{line:invokestv}), when a candidate exceeds the quota (line~\ref{line:quota-exceed}), their excess votes are transferred to the next eligible preferences of the ballots that contain the candidate (line~\ref{line:tranfer-excess}). In each round of ballot iteration, if no candidate has reached the quota, the candidate with the least vote(s) is eliminated (line~\ref{line:eliminate}). This candidates' excess votes are transferred to the next eligible preference of the ballots that contain the candidate that received the least votes (line \ref{line:transferelimnate}). The elimination of candidates stops when no more candidates can be eliminated to achieve the committee size (line~\ref{line:stopelim}). At this point, even though the remaining candidates did not receive enough votes to reach the quota, they are elected as part of the committee (line~\ref{line:elect}).

\subsection{Proofs of secure governance}

In this section, we show that BFT-STV (Algorithms~\ref{alg:stv} and~\ref{alg:stv1}) solves the secure governance problem (Def.~\ref{def:governance}).
To this end, the first theorem shows that the BFT-STV protocol ensures \emph{Proportionality}.
As mentioned in~\cref{ssec:gov}, recall that $n$, $m$ and $k$ denote the number of voting governors, the number of candidates and the targeted committee size, respectively.
As we consider byzantine nodes, note that the proof holds even if malicious voters vote in the worst possible way (e.g., based on what the others have voted). 

\begin{theorem}
	\label{theorem1}
	The BFT-STV multi-winner election protocol satisfies Proportionality.
\end{theorem}

\begin{proof}%
By examination of the code of Algorithms~\ref{alg:stv} and \ref{alg:stv1}, the only difference between BFT-STV and STV is the number of votes needed to elect a candidate. STV typically starts with $n$ received ballots 
whereas the BFT-STV starts the election as soon as $(n-t)$ ballots are received (line~\ref{line:nmtballots} of Alg.~\ref{alg:stv}), where $t$ is the upper bound on the number $f$ of byzantine nodes and $n$ is the total number of governors eligible to vote. This number of BFT-STV ballots is  distributed among a larger number of candidates. This can result in less than $k$ candidates receiving enough votes to reach the classic STV quota where $k$ is the size of the committee. 
By the Proportionality definition (Def.\cref{ssec:gov}), we need to show that if $j\cdot(n-t)/(k+1)$ voters put the same $s$ candidates as the top $s$ candidates in their ballot preference,
then those $s$ candidates will still be elected.
The proof follows from~\cite[p. 48--49]{janson2018thresholds}:
line~\ref{line:eliminate} of Algorithm \ref{alg:stv1} indicates that by elimination, the votes will still be concentrated on the top $j$ candidates such that  $0 < j \leq s$.
As a result, $j$ of those $s$ candidates will still be elected satisfying Proportionality.
\end{proof}

The next theorem shows that the BFT-STV protocol ensures \emph{Non-dictatorship} as defined in Definition~\ref{def:governance}. 
\begin{theorem}
	\label{thm:non-dictatorship}
	The BFT-STV multi-winner election protocol satisfies Non-dicatorship.
\end{theorem}
\begin{proof}

The proof shows the existence of an input of correct nodes for which 
a single adversary controlling $f$ byzantine nodes cannot have its preference $\ms{b}_a$ be the winning committee. 
Let $\ms{b}_a[-1]$ be the least preferred candidate of the adversary, we show that there exist preferences $b_1, ..., b_{n-f}$ from correct nodes such that the winning committee includes $b_a[-1]$. The result then follows from the assumption $k < m$.

By examination of the pseudocode, the winning committee is created only after receiving $n-t$ correctly formatted ballots (line~\ref{line:wellformed} of Alg.~\ref{alg:stv}).
By assumption, there can only be at most $f\leq t < n/3$ ballots cast by byzantine nodes. As a result, among all the $n-t$ received ballots, there are at least $n-2t > n/3$ ballots cast from correct nodes. In any execution, an adversary controlling all the byzantine nodes could have at most $f$ ballots as the adversary cannot control the ballot cast by correct nodes.
Let $b_1, ..., b_{n-f}$ be the ballots input by correct nodes to the protocol such that their first preference 
is the least preferred candidate of the adversary, i.e., $\forall i \in \{1, n-t\} : b_i = \ms{b}_a[-1]$.
Because $f\leq t < n/3$, we know that $\ms{b}_a[-1]$ will gain more votes than any of the other candidates, and will thus be the first to be elected (line~\ref{line:quota-exceed} of Alg.~\ref{alg:stv1}). By assumption, we have $k<m$, which means that there is a candidate 
the adversary prefers over  $b_a[-1]$ that will not be part of the winning committee.
Hence, this shows the existence of an execution where despite having an adversary controlling 
$f$ byzantine nodes, the adversary preference is not the winning committee.
\end{proof}

\remove{This variant removes the candidates with the least votes if the quota is not reached when all cast votes have been counted. The candidates that remain when all are eliminated, despite not reaching the quota are added to the committee.
At the end, the function returns the public keys and IP addresses of  the $1^{st}, 2^{nd}, ..., k^{th}$ ranked nodes.}

\section{\solution: Enabling Blockchain Social Choice}\label{sec:solution}

In this section, we present the blockchain called \solution and how it provides the swift governance reconfiguration based on our BFT-STV smart contract (\cref{sec:stv}). We show that \solution solves the Blockchain problem (Def.~\ref{def:blockchain}) in \cref{ssec:bc}.
The design of \solution is inspired by the open permissioned Red Belly Blockchain~\cite{CNG21}: while any client can issue transactions without permission, a dynamic set of permissioned consensus participants decide upon each block.
As a result, \solution ensures instant finality (by not forking), and is optimally resilient in that it tolerates any number $f\leq t<n/3$ of byzantine (or corrupted) nodes. However, \solution differs from the Red Belly Blockchain by mitigating bribery attacks and by integrating the Ethereum Virtual Machine (EVM)~\cite{Woo15} to support smart contracts necessary to offer the swift proportional reconfiguration.

\subsection{Reconfigurable governance with the BFT-STV smart contract}\label{ssec:reconf}


We now present how the blockchain is reconfigured with the new consensus committee once the BFT-STV smart contract elects the committee.
Offering proportional representation and non-dictatorship is not sufficient to cope with an adaptive adversary.
In order to mitigate bribery attacks, we now propose a swift reconfiguration that complements the BFT-STV algorithm.
Provided that you have $n$ nodes, it is sufficient to have $n/3$ corrupted nodes among them to make the consensus service inconsistent. This is because consensus cannot be solved with $n/3$ byzantine nodes~\cite{LSP82}.
In particular, it is well-known that neither safety (agreement) nor liveness (termination) of the consensus can no longer be guaranteed as soon as
the number of failures reaches $n/3$~\cite{CGG21}.
As a result, a coalition of $n/3$ byzantine nodes can lead the set of nodes to a disagreement about the next block to be appended to the chain.
An attacker can leverage these conflicting blocks in order to double spend: for example if it has two conflicting transactions in those blocks.

\subsubsection{How to ensure the existence of candidates}\label{ssec:bootstrap}

In order to bootstrap, an initial set of candidate nodes willing to provide the blockchain service and voter nodes is provided as part of the blockchain instance. 
Upon each block creation and as in classic blockchain solutions~\cite{Nak08,Woo15}, \solution offers a reward to each voter.
We assume that this reward incentivizes sufficiently many nodes to be candidates at all times.
There are few restrictions that need to be enforced in order to guarantee that the elected committee will proportionally represent voter nodes. First, the voters and the candidates set has to be large enough to represent all groups to which blockchain users belong. Second, the voters should not try to elect themselves---this is why we restrict the set of candidates to be disjoint from the set of voters, as we explained in \cref{ssec:bftstv}, Fig.~\ref{fig:bftstv} and as in other blockchains~\cite{BCC20,EOS-DPOS}.
These candidates and voters are typically encoded in the blockchain when the blockchain instance is launched, either by hardcoding their public key and network identifier (e.g., domain name or static IP addresses) in the genesis block or by invoking a default smart contract function at start time that records this information.
The voters then participate in the election: they cast their ballot by calling another function of the smart contract and passing it a list of candidates ranked in the order of their preferences. A ballot contains the network information of candidates in the order that the voter prefers. (We will present an implementation using static IP addresses in~\cref{ssec:setup}.)


%

\begin{algorithm}[h!]
	\caption{Reconfiguration of consensus service at a blockchain node\label{alg:recon}}
	{\footnotesize
	\begin{algorithmic}[1]
	
		\Part{upon receiving committee $S$} \Comment{smart contract emits event at Alg.\ref{alg:stv}, line~\ref{line:emit}}\label{line:emitconsenus}
		
		\State $\lit{stop}(\lit{\textsc{consensus-service}})$\Comment{stop the consensus service}\label{line:stopconsensus}
		\ForAll{$ip \in S$}\Comment{for IPs in committee}
				\State $A \gets A \cup \{\ms{ip}\}$ \Comment{add the IP address}
		\EndFor
		\If{$\ms{my-ip} \in A$}\Comment{if consensus node's IP is in the selected committee}
				\State $\lit{configure}(A)$\Comment{reconfigure node with the new committee peers}\label{line:configure}
				
				\State $\lit{start}(\lit{\textsc{consensus-service}})$\Comment{Start the consensus service}\label{line:startconsensus}
				\State $\lit{udpdate}(\lit{\textsc{dns-service}})$ \Comment{to redirect clients to the blockchain service}
		\EndIf
		\EndPart

	\end{algorithmic}%
}%
\end{algorithm}%

\subsubsection{Reconfiguration}

In this section, we present the reconfiguration (Algorithm~\ref{alg:recon}) of the blockchain that allows switching from the current committee to the new committee $S$ elected with the BFT-STV (\cref{sec:stv}) smart contract.
Once the network identifiers (domain names or static IP addresses) of the newly selected committee $S$ of governors is emitted by the BFT-STV protocol (line~\ref{line:emit} of Algorithm~\ref{alg:stv}), 
the blockchain nodes are notified with a smart contract event (line \ref{line:emitconsenus} of Algorithm \ref{alg:recon}).
The reception of this smart contract event triggers the stopping of the consensus service in the blockchain node through a {\ttt web3js} code (line~\ref{line:stopconsensus} of Algorithm~\ref{alg:recon}). 
After the consensus service is stopped, the elected consensus services are reconfigured with the network identifiers of the newly elected nodes (line~\ref{line:configure} of Algorithm~\ref{alg:recon}). Finally, the blockchain consensus service is restarted (line~\ref{line:startconsensus} of Algorithm~\ref{alg:recon}) to take this new committee into account.

\subsection{The transaction lifecycle}\label{sec:lifecycle}

In the following, we use the term \emph{transaction} to indistinguishably refer to a simple asset transfer, the upload of a smart contract or the invocation of a smart contract function. 
We consider a single instance of \solution whose genesis block is denoted by $B_0$ and whose blockchain nodes offer a consensus service depicted in Algorithm~\ref{alg:consensus} and a state service depicted in Algorithm~\ref{alg:sevmexecution}.
Although \solution also includes smart contracts, its transaction lifecycle is similar to~\cite{CNG21} and goes through these subsequent stages: 
\begin{itemize}
\setlength{\itemindent}{.56in}
\item[{\bf 1. Reception.}] The client creates a properly signed transaction and sends it to at least one \solution node. 
Once a request containing the signed transaction is received (line~\ref{line:reception} of Algorithm~\ref{alg:selection})
by the JSON RPC server of our blockchain node running within \solution, the validation 
starts (line~\ref{line:isvalid1} of Algorithm~\ref{alg:selection}).
If the validation fails, the transaction is discarded. If the validation is successful, the transaction is added to the mempool 
(line~\ref{line:addmempool} of Algorithm~\ref{alg:selection}).

\begin{algorithm}[h!]
	\caption{Selection of new transactions to decide\label{alg:selection}}
	{\footnotesize
	\begin{algorithmic}[1]
	

		\Part{$\lit{receive}(\lit{write}, \ms{tx})$}\Comment{State node upon receiving a transaction} \label{line:reception}
		\If{$\lit{is-valid}$} \Comment{if tx is validated}\label{line:isvalid1}
		  	\State $\ms{mempool} \gets \ms{mempool} \cup \{\ms{tx}\}$  \Comment{add it to mempool} \label{line:addmempool}
		\EndIf

		\WUntil{|$\ms{mempool}$| = $\lit{threshold}$ or $\ms{timer}$ expired}\Comment{wait sufficiently}\label{line:waittimer}\EndWUntil  
		
		\ForEach{$\ms{tx}$ in $\lit{mempool}$} 
		\State $\ms{prop.txs} \gets \ms{prop.txs} \cup \{\ms{tx}\}$ \Comment{create proposal} \label{line:createprops}
		\State $\ms{mempool} \gets \ms{mempool} \setminus \{\ms{tx}\}$ \Comment{remove tx from mempool}
		\EndFor
		\State $\ms{prop.timestamp} \gets \ms{timestamp}$ \Comment{add timestamp to proposal}\label{line:timestamp}

		\State $\ms{superblock} \gets \lit{propose}(\ms{prop})$ \Comment{propose to consensus, return superblock}\label{line:toconsensus}
        
       		\State $\lit{exec}(\ms{superblock})$ \Comment{execute superblock}

		\EndPart
		
		\algstore{alg:blockchain1}
	
	\end{algorithmic}%
}%
\end{algorithm}%

If the number of transactions in the mempool reaches a threshold of transactions or a timer has expired (line~\ref{line:waittimer} of Algorithm~\ref{alg:selection}), 
then the blockchain node creates a proposal of transactions with the transactions in the mempool (line~\ref{line:createprops} of Algorithm~\ref{alg:selection}). Consequently, a timestamp is added to the proposal (line~\ref{line:timestamp} of Algorithm~\ref{alg:selection}) and proposed to the consensus service (line~\ref{line:toconsensus} of Algorithm~\ref{alg:selection}). 
\item[{\bf 2. Consensus.}] As in the Democratic Byzantine Fault Tolerant (DBFT)~\cite{CGLR18}, upon reception of the proposal, the blockchain node reliably broadcasts the proposal to other blockchain nodes (line~\ref{line:rbbroadcast} of Algorithm~\ref{alg:consensus}) in the same consensus instance. Lines~\ref{line:dbft1}--\ref{line:dbft2} of Algorithm~\ref{alg:consensus} present the section of our consensus algorithm. 
We point the reader to~\cite{CNG21} for a detailed description of the consensus protocol and to~\cite{BGK21} for the formal verification of its binary consensus. In short, the consensus protocol waits until all proposals have been received for binary consensus instances that have decided 1 (line~\ref{line:wttillprops} of Algorithm~\ref{alg:consensus}) and forms a superblock out of those proposals (line~\ref{line:createsupblock} of Algorithm~\ref{alg:consensus}). Finally, the consensus service delivers the superblock to the state service (line~\ref{line:toconsensus} of Algorithm~\ref{alg:selection}). 

\begin{algorithm}[h!]
	\caption{Consensus protocol\label{alg:consensus}}
	{\footnotesize
	\begin{algorithmic}[1]

	\algrestore{alg:blockchain1}

		\Part{$\lit{propose}(\ms{prop})$}\Comment{starting the consensus algorithm}
		\State $\lit{reliable-broadcast}(\ms{prop}) \to \ms{props[k]}$ \Comment{reliably deliver in $\ms{props}$ array}\label{line:rbbroadcast}
		\While{$|\{k : \ms{bitmask}[k]=1\}|< n-t$ or ${timer}$ did not expire}\label{line:dbft1}
				\State {\bf for all}~{$k$ such that $\ms{props}[k]$ has been delivered}
				\State \T$\ms{bitmask}[k] \gets \lit{bin-propose}_k(1)$\Comment{input 1 to binary consensus~\cite{CGLR18}}\label{line:bin-cons1}
        \EndWhile
        \State {\bf for all}~{$k$ such that $\ms{props}[k]$ has not been delivered}
        \State \T $\ms{bitmask}[k] \gets \lit{bin-propose}_k(0)$ \Comment{propose 0 to $k^{\ms{th}}$ binary consensus}\label{line:dbft2}
        \WUntil{$\ms{bitmask}$ is full and $\forall \ell,\ms{bitmask}[\ell] = 1 : \ms{props}[\ell] \neq \emptyset$}\EndWUntil\label{line:wttillprops}  
        \State $\forall \ell\text{ s.t. }\ms{props}[\ell] \neq \emptyset :$\label{line:ref}
         $\ms{superblock} \gets 
         \ms{bitmask}  ~\&~ \ms{props}$ \label{line:createsupblock}\Comment{bitwise \&}
        \Return $\ms{superblock}$ \EndReturn 
		\EndPart

	\algstore{alg:blockchain2}
	
	\end{algorithmic}%
	}%
\end{algorithm}%

\remove{
The consensus system starts a new instance of consensus using the new block if it is not currently part of another consensus instance. Otherwise, it adds the new block to the block queue waiting for the current consensus instance to terminate. The consensus execution 
consists of  an all-to-all reliable broadcast of the blocks among all consensus replicas that trigger as many \emph{binary} consensus instances whose outputs indicate the indices of acceptable blocks 
as detailed in~\cref{sec:dbft}. The consensus system creates a superblock as in~\cite{CNG21} with all acceptable blocks and sends this superblock to the state machine.}
\setlength{\itemindent}{.43in}
\item[{\bf 3. Commit.}] Once the superblock is received by the blockchain node (Algorithm~\ref{alg:selection}, line~\ref{line:toconsensus}) the commit phase starts. Firstly, each proposal is taken in-order from the superblock and each transaction in it is validated, i.e., its nonce and signature are checked as correct~\cite{Woo15} (Algorithm~\ref{alg:sevmexecution}, line~\ref{line:valid2}). If a transaction is invalid it is discarded. If a transaction is valid, the transaction is executed and the EVM state trie is updated (Alg.~\ref{alg:sevmexecution}, line~\ref{line:updatestate}). The execution of a transaction returns the updated state trie $\ms{S_{next_{k}}}$ and a transaction $\ms{receipt}$ (Alg.~\ref{alg:sevmexecution}, line~\ref{line:updatestate}). All the executed transactions in a proposal are written to the transaction trie (Alg.~\ref{alg:sevmexecution}, line~\ref{line:start-var}) and all the receipts are written to the transaction receipt trie (Alg.~\ref{alg:sevmexecution}, line~\ref{line:receipttrie}). A block is constructed (Alg.~\ref{alg:sevmexecution}, line~\ref{line:end-var}) with hashes returned by the $h$ function, the gas used $\ms{GU}$ by all transactions and the gas limit $\ms{GL}$ that can be consumed during the block execution (Alg.~\ref{alg:sevmexecution}, line~\ref{line:hashes}), as in Ethereum~\cite{Woo15}. Finally the block is appended (Alg.~\ref{alg:sevmexecution}, line~\ref{line:appendblock}) to the blockchain of the blockchain node.
\remove{Once the superblock is decided, each of its blocks are executed, their hash is included, their results are written to persistent storage on the local disk and the lifecycle ends.}

\begin{algorithm}[h!]
	\caption{Execution of a superblock by a blockchain node\label{alg:sevmexecution}} 
	{\footnotesize
	\begin{algorithmic}[1]
	
	\algrestore{alg:blockchain2}

		\Part{$\lit{exec}(\ms{superblock})$}\Comment{parse superblock to the SEVM}\label{line:execution}
	    \State $\ell \gets 1$ 
	    \ForEach{$k \in [0..] : \ms{props[k]}$ in 
	    $\ms{superblock}$}\label{line:props} \Comment{order transactions...}
	     \ForEach{$tx$ in $\ms{props[k].txs}$} \Comment{...to minimize conflicts}
	     		\If{$\lit{is-valid}(tx)$} \Comment{if validated}\label{line:valid2}
				\State $\tup{\ms{S_{next_{k}}}, \ms{receipt}} \gets \lit{run}(\ms{tx}, \ms{S})$ \Comment{run tx, return state and receipt}\label{line:updatestate}
	     			\State$\ms{receipts_{k}} \gets \ms{receipts_{k}} \cup \{receipt\}$\Comment{collect all receipts}
	     			\State$\ms{val_{k}} \gets \ms{val_{k}} \cup \{tx\}$\Comment{collect executed txs from $\ms{props[k]}$ to $\ms{val_{k}}$}\label{line:validity1}
	     		\EndIf
	     \EndFor
	     \State $\ms{timestamp}_{k} \gets \ms{props[k].timestamp}$ \Comment{the $k^{\ms{th}}$ timestamp}
	     \State $\ms{TX}_{k}$ $\gets$\label{line:start-var} $\lit{update-tx-trie}(\ms{val}_{k})$\Comment{update transaction trie}
	     \State$R_{k}$ $\gets$ $\lit{update-receipt-trie}(receipts_{k})$\Comment{update transaction receipts}\label{line:receipttrie}\label{line:assign-rk}
	     \State$B_{\ell}$ $\gets \langle h(\lit{header}(B_{\ell-1})), h(\ms{S_{next_{k}}}), h(\ms{TX}_{k}), h(R_{k}), \ms{GU}_{k}, \ms{GL},$\label{line:hashes}
	     \State \T\T\T$\ms{nonce}, \ms{timestamp}_{k}, \ms{val}_{k}\rangle $ \Comment{create block}\label{line:end-var}
	     
	     \State $\ms{chain} \gets \ms{chain} \cup \{B_{\ell}\}$\Comment{append block to blockchain}\label{line:appendblock}
	     \State $\ell \gets \ell +1$
	    \EndFor
		\EndPart

	\end{algorithmic}%
}%
\end{algorithm}%
\end{itemize}

\remove{

\subsection{From the EVM to SEVM}\label{sec:sevm}

Here we present the modifications we made to the original EVM (and in particular {\ttt geth} v1.8.27) in order to obtain the Scalable EVM, or SEVM. 
In particular, we reduce the average number $k$ of transactions each {\ttt geth} server eagerly validates in a system receiving $k$ transactions to $k/n$ in the good case.

\paragraph{Reducing the transaction validations}
As opposed to each Ethereum server that validates 
eagerly and lazily each of the $k$ transactions of the system---be it a native payment, a smart contract creation or its invocation---each of the $n$ \solution servers eagerly validates on average $k/n$ transactions.
Specifically, only one SEVM node  needs to eagerly validates each transaction: the first SEVM node receiving the transaction validates it but does not propagate it to other SEVM nodes. 
As a result, \solution limits the redundant validations, which improves performance. More precisely, if the number of SEVM nodes is $n$, then each SEVM node does $1+1/n$ validations per transaction on average (one lazy validation + $1/n$ eager validation) compared to the two validations needed in {\ttt geth}. As $n$ tends to infinity, \solution servers validate on average half what {\ttt geth} servers validate.
In the worst case, where all clients send their transactions to $f+1 = n/3$ servers simultaneously, then each server will still eagerly validate only $k/3$ transactions. This helps ensuring the \emph{close-to-second-responsiveness} property (\cref{sec:bgrd}).

\paragraph{Reliably storing superblocks}
As our consensus aims at outputting $n$ blocks instead of 1, our SEVM must store multiple blocks in a row.
In blockchains, blocks are typically created at the beginning of the consensus. This is why {\ttt geth} expects to invoke the {\ttt updateBlockState} function, which sets the state parameters of a block, only once per consensus instance:
{\ttt updateBlockState} points a block to its parent, assigns the block header timestamp, block difficulty and the number of and transaction receipts associated with a block. 
This is why, this function should be invoked for each block before the transactions of the block are executed and persisted, in order to ensure that the tries are updated properly.
To store multiple blocks at the end of the consensus intance, 
we modified {\ttt geth} to invoke {\ttt updateBlockState}  multiple times 
per consensus instance (one invocation per block) as follows:


More specifically, we changed the original procedure 
to guarantee that not only the last block but all preceding ones in our superblock were correctly stored in the transaction and reception tries as a batch of $n$ blocks.  
Like the C++, python and {\ttt geth} software of Ethereum, 
we reliably store the information in the open source key-value store LevelDB.

\paragraph{SEVM support for fast-paced consecutive blocks}
Since our consensus system is fast, it creates and delivers superblocks at high frequency through the commit channel to the SEVM. As {\ttt geth} does not expect to receive blocks at such a high frequency, it would raise an exception outlining that consecutive block timestamps are identical, which never happens in a normal execution of Ethereum.
This equality arose because  {\ttt geth} encodes the timestamp of each block as {\ttt uint64}, not leaving enough space for encoding time with sufficient precision.
We changed the original check that compares the parent block timestamp to the current block timestamp  and returns the zero block time error in {\ttt go-ethereum/consensus/ethash/consensus.go} when {\ttt header.Time <= parent}: SEVM reports an error only when the difference is strict, i.e., {\ttt header.Time < parent.Time}. \vincent{This should appear in the pseudocode.}

\paragraph{Bypassing the SEVM resource bottlenecks}
After the consensus, the SEVM lazily validates many transactions, updates the memory (state trie) and storage (reception/transaction tries), which consume CPU, memory and IO resources, respectively.
Typically, high CPU usage slows down the SEVM and when the RAM is overused, swap space in slower media like SSD or disks are used, which slows down memory reads from disks resulting in a bottleneck. 
Due to our superblock optimizations, we observed that consuming each resource one after another, for 20 proposed blocks with a total of 30,000 transactions, would produce an out-of-memory (OOM) kill even on our reasonably-provisioned AWS instances featuring 16\,GB RAM and 4 vCPUs (cf.~\cref{fig:superblock}).
This is why we made SEVM fully process one proposed block of the superblock at a time
allowing it to alternate frequently between CPU-intensive (verifying signatures), memory-intensive (state trie write) and IO-intensive (reception/transaction tries writes) tasks.
Thanks to alternating, SEVM does not experience bottlenecks as the number of nodes increases (cf.~\cref{sec:scalability}).

} 

\subsection{A BFT consensus for non-dictatorship}\label{sec:dbft}
As \solution builds upon DBFT~\cite{CGLR18} it is inherently democratic.
In fact, DBFT is leaderless which favors non-dictatorship as we explain below: as indicated in Algorithm~\ref{alg:consensus}, DBFT reduces the problem of multi-value consensus (to decide upon arbitrary values) to the problem of binary consensus (to decide upon binary values exclusively) by executing an all-to-all reliable broadcast algorithms (line~\ref{line:rbbroadcast}) followed by 
up to $n$ binary consensus instances (lines~\ref{line:bin-cons1} and \ref{line:dbft2}) running in parallel. 
As one can see, in Algorithm~\ref{alg:consensus} all nodes execute the same code and none plays the role of a leader.

By contrast, classic BFT algorithms~\cite{CL02,BSA14,Kwo14,BKM18,GAG19,YMR19} used in blockchains are generally leader-based. 
They proceed as follows: A special node among $n$, called the \emph{leader}, sends its block to the rest of the system for other nodes to agree that this block is the unique decision. 
If the leader is slow, then another leader is elected but eventually the block of some leader is decided by all. 
The drawback with such solutions is that the leader can be byzantine and thus propose the block with the content it chooses. 
This limitation of leader-based consensus algorithms is precisely the motivation for a recent work~\cite{ZSC20} that aims at circumventing byzantine oligarchy by ensuring that if non-concurrent transactions are perceived in some order by all correct nodes, then they cannot be committed in a different order.

\begin{figure}[t]
	\centering
	\subfigure[First, the byzantine leader (in the middle) receives transaction ballots from correct nodes\label{sfig:leader1}]{\includegraphics[scale=0.55]{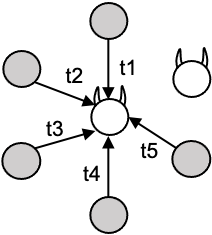}}\hspace{1.6em}
	\subfigure[Then, the byzantine leader receives the transaction ballots from the byzantine nodes\label{sfig:leader2}]{\includegraphics[scale=0.55]{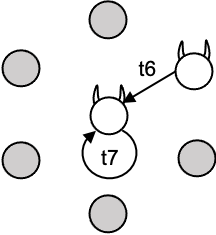}}\hspace{1.6em}
	\subfigure[Finally, the byzantine leader imposes the block with $n-t$ ballots including ballots from the byzantine nodes to other nodes, hence favoring byzantine preferences\label{sfig:leader3}]{\includegraphics[scale=0.55]{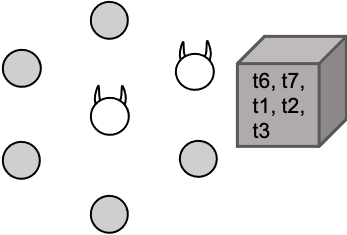}}
	\vspace{-1em}
	\caption{A blockchain running a leader-based consensus with $n=7$ and $f=2$ may end up favoring byzantine preferences during an election if the leader is byzantine}
	\label{fig:leader}
\end{figure}

To place this problem in our context, consider Figure~\ref{fig:leader} depicting an execution of BFT-STV (Algorithm~\ref{alg:stv}) in one of these leader-based blockchains with $n=7$ and $f=t=2$.
(Recall that without assuming synchrony, the smart contract cannot wait for $n$ transactions from distinct processes as it cannot distinguish between a slow process and a byzantine process that do not send its invocation. As a result the smart contract can only wait for at most $n-t$ invocations before continuing its execution to avoid blocking forever.)
Consider that in this particular execution, the $n-t$ correct nodes issue a transaction (Fig.\ref{sfig:leader1}) invoking the function $\lit{cast-ballot}$ with their candidate preferences before any byzantine node issued any transaction invoking $\lit{cast-ballot}$ (Fig.\ref{sfig:leader2}). These transactions are not yet committed and are still pending, when the leader creates its block by including the transactions it has received. If the leader is byzantine, it may wait and gather transactions issued by all the byzantine nodes of its coalition before creating its proposed block. Once it receives $t$ transactions from byzantine nodes it then creates its block with $t$ transactions issued from byzantine nodes and $n-2t$ transactions issued by correct nodes (Fig.\ref{sfig:leader3}), before proposing it to the consensus. Once this block is decided, the transactions it contains that invoke $\lit{cast-ballot}$ are executed. As a result, the BFT-STV smart contract finally returns a committee voted upon by $t$ byzantine nodes and $n-2t$ correct nodes, whereas without the leader the BFT-STV smart contract could have returned a committee voted upon by $n-t$ correct nodes.

\solution is immune to this dictatorship problem without assuming synchrony due to its democratic consensus algorithm~\cite{CGLR18}. More precisely, Algorithm~\ref{alg:consensus} does not decide a single block proposed by any particular node or leader. Instead it decides the combination of the blocks proposed by all consensus participants, hence called \emph{superblock} (line~\ref{line:createsupblock}) as in~\cite{CNG21}. All nodes reliably broadcast its block and participate in a $k^{th}$ binary consensus by inputting value 1 if it received the proposal from the $k^{th}$ node. If no block is received from the $k^{th}$ node, then the $k^{th}$ binary consensus is invoked with input value 0. The outcome of these $n$ binary consensus instances are stored in a bitmask (lines~\ref{line:bin-cons1} and \ref{line:dbft2}).  Finally the bitmask is applied to the array of received blocks to extract all transactions that will be included in the superblock (line~\ref{line:createsupblock}). Hence the inclusion of a particular block is independent of the will of a single node, that could otherwise act as a dictator.

\remove{The protocol is divided in two procedures, {\ttt start\_new\_consensus} that spawns a new instance of (multivalue) consensus by incrementing the replicated state machine {\ttt index}, and {\ttt propose} that ensures that the consensus participants find an agreement on a superblock comprising all the proposed blocks that are acceptable. 
The execution of {\ttt propose} follows a classic reduction of the multivalue consensus problem to the binary consensus problem~\cite{BCG93,BKR94,CGLR18,CGG19}. It thus starts by executing a reliable broadcast~\cite{B87}, which guarantees that any block delivered to a correct process is delivered to all the correct processes.
For each value reliably delivered, the process spawns the DBFT binary consensus (omitted here by lack of space)~\cite{CGLR18} and formally proved correct~\cite{TG19} with proposed value 1.}

\remove{As soon as some of these binary consensus instances return 1, the protocol spawns binary consensus instances with proposed value {\ttt false} for each of the non reliably delivered blocks. 
Note that as the reliable broadcast fills the {\ttt block} in parallel, it is likely that the blocks reliably broadcast by correct processes have been reliably delivered resulting in as many invocations of the binary consensus with value {\ttt true} instead. Once all the $n$ binary consensus instances have terminated, i.e., {\ttt decidedCount == n}, the superblock is generated with all the reliably delivered blocks for which the corresponding binary consensus returned {\ttt true}.
At the end of {\ttt start\_new\_consensus}, if the superblock of the consensus contains the block proposed, then this block is removed from the {\ttt blockQueue} to avoid reproposing it later.}


\section{Evaluation of \solution}\label{sec:evaluation}\label{ssec:setup}

In this section, we evaluate the performance of \solution on up to 100 machines located in 10 countries. 
We measure the performance of reconfiguring the blockchain nodes in SocChain in terms of the time taken to stop the blockchain service with the previous committee and re-start the blockchain service but with a new committee. We also
evaluate the execution time of the BFT-STV smart contract depending on the number of candidates and voters. 
Finally, we show that a growing number of blockchain participants marginally impacts \solution's performance.


\subsection{Experimental settings}

We now present the experimental settings of our evaluations dedicated to test the performance.


\subsubsection{Controlled setting}
In order to evaluate \solution, we deployed \solution on up to 100 VMs. 
In order to combine realistic results and some where we could control the network, we deployed VMs on Amazon Web Services (AWS) and an OpenStack cluster 
where we control the delay using the {\ttt tc} command. We also measured world-wide delays using 10 availability zones of AWS located across different countries.
On AWS, we deployed up to 100 VM instances with 8 vCPUs and 16\,GB of memory connected through Internet.
On OpenStack we deployed
up to 20 VM instances with 8 vCPUs and 16\,GB of memory connected through a 10\,Gbps network where
we added artificial network delays under our control to represent the latencies observed over the Internet across different geographical regions.

\subsubsection{Client setup}
In our experiments, we send a fixed number of transactions from each client machine to keep the sending rate constant. 
Each client instance sends a distinct set of transactions to a specific blockchain node. The sending is done concurrently to each blockchain node so as to stress-test the blockchain.  

\remove{

\paragraph{Comparison with other blockchains.}
As baselines, we evaluated two blockchains that both have support for the Ethereum smart contracts and builds upon a BFT consensus algorithm.
\begin{smallitem}
\item {\bf Quorum}~\cite{jpmorganchase_quorum} is an open source blockchain developed by JP Morgan and acquired by ConsenSys that  supports Ethereum smart contracts. It builds upon {\ttt geth} and IBFT that has recently been corrected to ensure termination~\cite{Sal19}. Its code is written in golang and is publicly available at {\ttt \url{https://github.com/ConsenSys/quorum}}. We evaluate quorum by setting it up as per~\cite{quorumsetup}.
\item {\bf Concord}~\cite{concord} is described as a scalable decentralized blockchain written in C++ that builds upon a C++ EVM implementation and the SBFT consensus algorithm~\cite{GAG19} that outperforms PBFT by reducing its message complexity using threshold signatures. The code of Concord is available at {\ttt \url{https://github.com/vmware/concord-bft/}}. While we could not find any performance result of its Concord blockchain, SBFT is known to achieve 172 transactions per second in a world-wide setting~\cite{GAG19}. 
\item {\bf Libra}~\cite{BBC19} We have tested Facebook Libra blockchain with its recommended most stable version from the testnet branch (\url{https://developers.diem.com/docs/core/my-first-transaction/}), which is available at \url{https://github.com/QSFTW/libra/tree/1529fa10155451e03a95837f9fe11dbc6ed379d4}. Unfortunately, the performance we obtained with non-conflicting transactions written in the Move programming language and sent from 4 distinct clients to 4 validator nodes could not exceed overall throughput was 10.5 TPS on average.
As a result, we do not report the results in our graphs.
\item {\bf Alternative blockchains} 
exist~\cite{burrow,Ethermint,GHM17},
however, we could not evaluate them. Burrow~\cite{burrow} has recently been reported as unstable~\cite{SNG20} while Ethermint~\cite{Ethermint} is in its pre-alpha development stage and we could not benchmark it due to some open issues\vincent{Verify that it is still accurate}. 
Finally, other blockchains claim to improve the performance of Ethereum but no publications describe their internals, like NEO, Tron and EOS.
Algorand~\cite{GHM17} requires nodes to access hardcoded DNS that limits its deployment in an evaluation testnet.
\end{smallitem}

\paragraph{Benchmarking with web3.js.}
We propose the following way to evaluate Ethereum blockchains using the javascript API, {\ttt web3.js}~\cite{DWC17,caliper,SNG20}:
\begin{smallitem}
\item {\bf Raw transaction method:} This method sends transactions to the SEVM node through {\ttt http}.
This is the method adopted in our experiments, except when specified otherwise, because we believe it to be the more realistic than the other.
We create wallets using {\ttt ethereumjs-wallet} and pre-signed transactions using {\ttt ethereumjs-tx} to offload the encryption time from the performance measurement. We serialize these transactions and save them to a JSON file. We iterate through 
this file 
and send the transactions to the EVM nodes  
using {\ttt web3.eth.sendSignedTransaction}. 
A generic implementation of the way in which we send transactions for our benchmarks looks as follows:
 \begin{lstlisting}[language=parameterized,basicstyle=\LSTfont,escapechar = ?,escapeinside={(*}{*)},frame = single]
  const w3 = require("web3")
  const fs = require("fs")
	
  const web3 = new w3(new w3.providers
    .HttpProvider("http://l0.16.18.254:8545"))
	
  // load the transaction file
  txs = JSON.parse(fs.readFileSync('txs.json'))
	
  // send a transaction one by one
  for ( let i = 0; i < txs.length; i++ ) {
    web3.eth.sendSignedTransaction("0x" + 
      txs[i]["serialized"])
      .on('receipt', console.log)
      .catch(console.log)
  }
\end{lstlisting}
By adjusting the loop iteration count, one can adjust the batch size. The {\ttt http} provider defines the IP address of the {\ttt geth} node. The port specifies the port that SEVM is running on.

\end{smallitem}

} 

\remove{

\subsection{Comparison With Other Blockchains}\label{sec:comparison}
We compare \solution to two Ethereum-compatible blockchains that build upon byzantine fault tolerant consensus algorithms, Quorum~\cite{jpmorganchase_quorum} and Concord~\cite{GAG19}. As none shares the optimizations of \solution, this gives an idea of their impact on performance.

\begin{figure}[t]
	\centering
	\includegraphics[scale=0.265,clip=true,viewport=55 0 960 290]{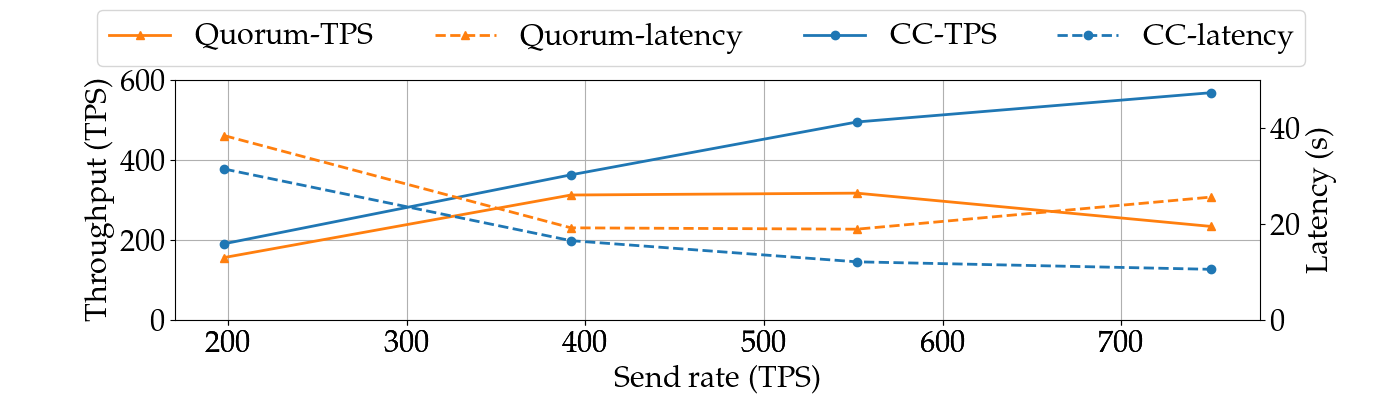}
	\caption{Comparison of throughput and latency between \solution (CC) and Quorum when sending 6000 transactions in total}
	\label{fig:quorum}
\end{figure}

\paragraph{Quorum vs \solution.}
For each point in these series of experiments, we computed the average throughput over 3 runs and increase the sending rate to observe the effect on the throughput.
Figure~\ref{fig:quorum} reports the latency and throughput of both \solution and Quorum.
We can see that the throughput of \solution is constantly higher than the throughput of Quorum and the latency of the former is constantly lower than the one of the latter. 
This is because of our superblock optimization used in the consensus as well as the changes made at the EVM to handle superblocks. The transaction validation reduction also seems to have improved throughput and reduce latency. 
Interestingly, we also observe that the performance of Quorum starts decreasing as the sending rate increases
whereas the performance of \solution keeps increasing, this is seemingly due to the growing backlog of Quorum that induces congestion.

\paragraph{Concord vs \solution.}

\deepal{Adding the new figure for comparison with concord by varying send rate}
At the time of writing, Concord suffers from known configuration issues~\cite{bugconcord20} that we could not resolve with the help of the development team.
As a result, we could not deploy Concord on multiple physical machines and could not initialize our own genesis block to create our own accounts and send pre-signed transactions. 
We confirmed with the development team that it was 
quite hard  to
deploy 
the publicly available code of Concord 
on multiple machines as this needed to be done manually.
To bypass this limitation, we experimented with web3.sendTransaction (\cref{ssec:setup}) to avoid signing transactions and performed the comparison on 4 docker container nodes (as recommended) 
hosted by the same physical machine. The physcial machine used for this experiment was an AWS ec2 c5.9xlarge instance to avoid the overhead of having multiple containers in the same machine.
There were 4 clients in total, each client sending 1500tx each to a total of 6000tx. 

\begin{figure}[t]
	\centering
	\includegraphics[scale=0.265]{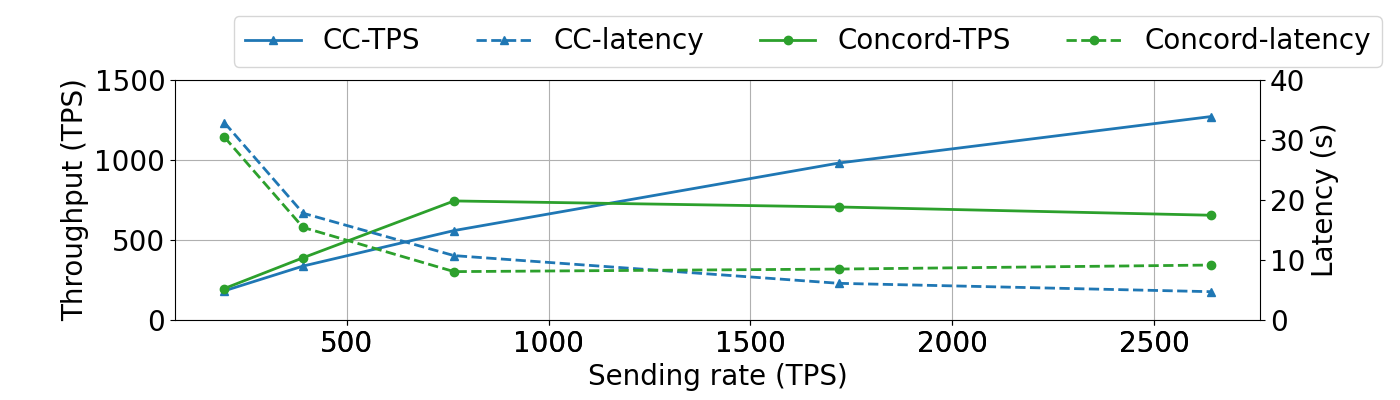}
	\caption{Throughput and Latency comparison of \solution (CC) and Concord against sending rate.\vincent{Can you center the legends?}}
	\label{fig:concord}
\end{figure}
Figure~\ref{fig:concord} compares the throughput and latency of Concord and \solution. 

First, we observe that the throughput of both blockchains is different from Figure~\ref{fig:quorum} 
but it is hard to compare because, as we discussed above, Concord could only run on a single physical machine with a different evaluation method so we had to do the same for \solution to make the comparison fair.

Second, we observe that the latency of concord is slightly lower and throughput of Concord is higher than \solution at lower sending rates. However, once the sending rate increases above ~1000TPS, the throughput of Concord starts to decline while the latency increases. On the other hand, \solution's throughput increases and its latency decreases when the sending rate increases. As seen by Figure~\ref{fig:concord}, the throughput of \solution is considerably higher than Concord above ~2500TPS of sending rate and continues to rise. The latency is also lesser than Concord at higher sending rates. This helps us conclude that in high traffic situations \solution performs much better and is more resilient.

\subsection{DApps and Intensive Benchmarks}\label{sec:dapps}

In this section we (i)~benchmark \solution with CPU-heavy and IO-heavy transactions, (ii)~evaluate it when running a banking and an e-voting DApps and (iii)~measure the impact of the validation reduction as well as the decoupling of the block storage (\cref{sec:sevm}) on performance. The resource-intensive smart contracts and the banking DApp, called ``smart bank'', were taken from BlockBench~\cite{DWC17,blockbench} whereas the e-voting DApp is from~\cite{e-voting}. We used 4 blockchain nodes for these evaluations. The CPU-heavy transactions are 6000 smart contract invocation transactions that quick-sort an unsigned array of length 128 in the worst case scenario, where the complexity is $\mathcal{O}(n^2)$. The IO-heavy transactions are 6000 smart contract invocation transactions that write key-value pairs to an array. However, we made a slight change in the CPU-heavy contract. Instead of passing a large array to the sort function, we stored the array in the smart contract itself, and sorted it once the sort function was called. The computations required to sort the array is still the same this way and has no bearing on the workload. Finally, the smart bank DApp essentially behaves as a bank on smart contract. It has functions for sending payments, updating balance and writing checks.

\begin{figure}[ht]
	\centering
	\includegraphics[scale=0.255,clip=true,viewport=3 0 930 150]{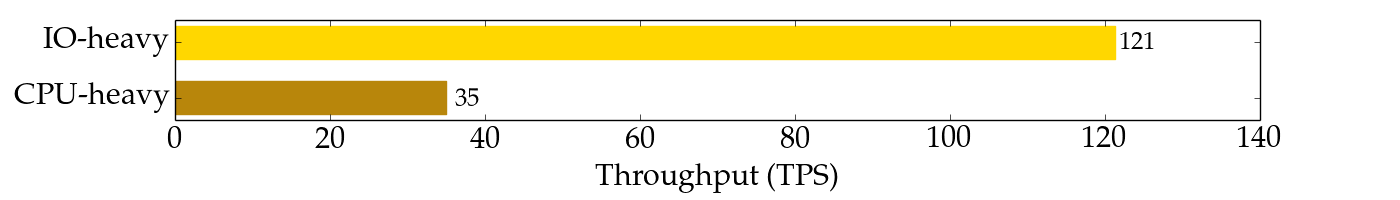}
	\caption{Throughput of \solution for CPU-heavy and IO-heavy functions\label{fig:CPU-IO}
}
\end{figure}

It must be emphasized that the low throughput values observed for CPU-heavy and IO-heavy transactions (Fig.~\ref{fig:CPU-IO}) are due to bottlenecking the transaction execution and requires unrealistically large gas limit. 
Interestingly, the performance under these extreme workloads is still better than Ethereum's maximum throughput mainly because of our consensus and superblock optimization.

In the smart bank DApp, we micro-benchmarked each function with 6000 transactions each. The results are shown in Figure~\ref{fig:smartbank}. The more realistic smart bank DApp peaks above 1000\,TPS, which shows 
that \solution treats realistic workloads at high speed.

\begin{figure}[ht]
	\centering
	\includegraphics[scale=0.255,clip=true,viewport=3 0 930 230]{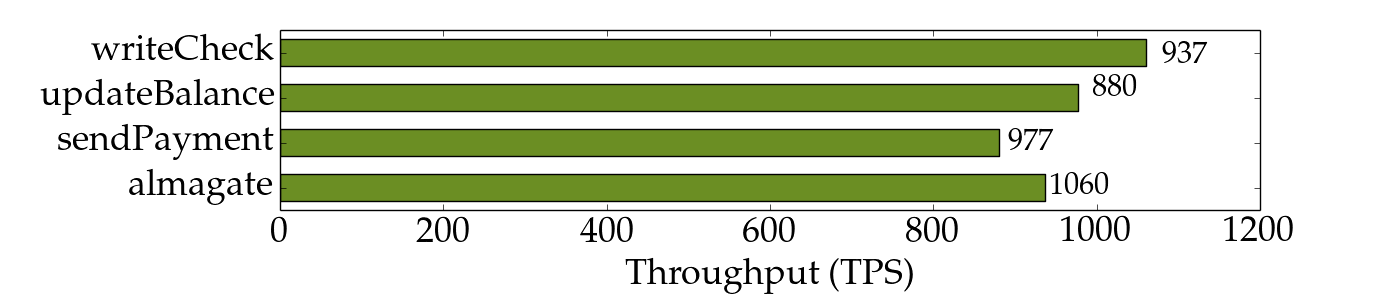}
	\caption{The throughput of \solution for each function of the smart bank DApp\label{fig:smartbank}}
\end{figure}

} 

\remove{ 
\deepal{Essentially we are saying the baseline is much faster?}
\subsection{Simple e-voting DApp baseline on \solution}\label{sec:evoting}
In this section, we measure the performance of a governance DApp, called e-voting~\cite{e-voting}, running on  \solution, as a baseline.  This type of governance DApps is becoming increasingly popular to automate the management of a decentralized governance. In particular, one can use such an e-voting function for adding and removing consensus participants.
This e-voting DApp runs on top of \solution, but as opposed to BFT-STV (\cref{sec:stv}), it allows users to cast votes about a single candidate.
As a result, this e-voting DApp cannot solve the secure governance problem (Def.\ref{def:governance}) because as it implements a single-winner voting system, it does not offer Proportionality. 

The purpose of this DApp is to cast a vote on a particular ballot. 
To this end, a ballot manager  deploys this DApp as a smart contract and adds voters with their account address and names using the {\ttt addVoter} function of the contract. The ballot manager then enables the votes by invoking the {\ttt startVote} function, which permits voters to cast a {\ttt true} or {\ttt false} vote using the {\ttt doVote} function. The ballot manager finally ends the vote by invoking the {\ttt endVote} function.

\begin{figure}[t]
	\centering
	\includegraphics[scale=0.28,clip=true,viewport=35 0 935 230]{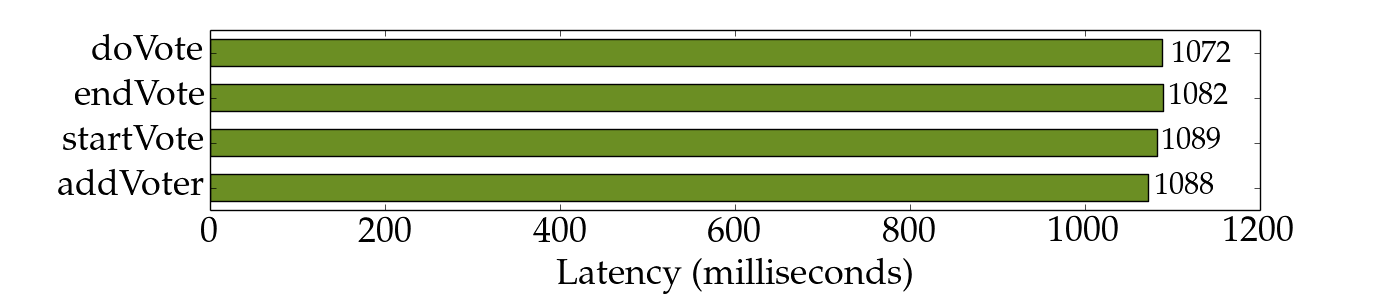}
	\vspace{-2em}
	\caption{The latency of \solution for each function of the e-voting DApp\label{fig:evoting}}
	\vspace{-1em}
\end{figure}

Figure~\ref{fig:evoting}  depicts the average latency of the e-voting DApp when deployed on four \solution nodes without additional delays.
In particular, this figure depicts the latency it takes to invoke each smart contract function, for the invocation transaction to be committed and returned. We observe that the end-to-end latency lies between 1 and 1.1 second for any function. First, we observe a low latency that is explained by the instant finality of our blockchain: as opposed to classic blockchains~\cite{Nak08}, any block appended to \solution is final because there are no forks.
For the sake of comparison, in Ethereum~\cite{Woo15} or Solana whose appended blocks can be changed later, the time it takes to commit a transaction is longer (about three minutes in expectation for Ethereum). Second, these results indicate that our implementation benefits from the lack of block periods. As the blocks are generated on-demand after the reception of new requests in \solution, there is no need to wait for a particular block period to commit. For the sake of comparison, Quorum~\cite{jpmorganchase_quorum} builds upon the Ethereum Virtual Machine (EVM) and a BFT consensus algorithm to also offer instant finality, but its default block period is 1 second hence the average time it would take Quorum to commit any transaction would necessarily be 1.5 second, 
which is larger than the latency of \solution. 

} 

\remove{

\paragraph{Effect of validation reduction.}

In order to assess the impact of the validation optimization (\cref{sec:sevm}) of the SEVM on the performance, 
we measured the time spent validating eagerly the smart bank DApp. To this end, 
we instrumented its {\ttt writeCheck} function to measure both the total time $\Delta^{n}_{SEVM}$ spent treating $k$ calls and the average time 
$\delta^{n}_{SEVM}$ spent by each server of SEVM validating eagerly these calls on $n$ nodes, 
to deduce the rest of the treatment time not affected by the validation optimization $\beta = \Delta^{n}_{SEVM} - \delta^{n}_{SEVM}$.

Based on this measurement, we could deduce the time $\delta_{EVM}$ the EVM would spend 
validating eagerly without the validation optimization:
$\delta_{EVM} = n\cdot \delta^{n}_{SEVM}.$
In particular, regardless of $n$, we know that the EVM would spend $\Delta_{EVM} = \beta + \delta_{EVM}$ to treat the function calls.
By contrast, depending on $n$, the SEVM would spend $\Delta^{n}_{SEVM} = \beta + \delta^{n}_{SEVM}$.
As $\delta^{n}_{SEVM} =  \delta_{EVM}/n$, we know that $\lim_{n\rightarrow \infty}(\delta^{n}_{SEVM}) = 0$.
This means that, with $n$ servers, 
the EVM slowdown compared to the SEVM is:
$$S = \frac{\Delta_{EVM} - \Delta^{n}_{SEVM}}{\Delta^{n}_{SEVM}} = \frac{\delta_{EVM}+\delta^{n}_{SEVM}}{\beta+ \delta^{n}_{SEVM}}.$$
As $n$ tends to infinity, we thus have a slowdown of: $$\lim_{n\rightarrow +\infty}S = \frac{\delta_{EVM}}{\beta}.$$

Our measurement obtained with $k=6000$ transactions and $n=4$ revealed that 
$\delta^{n}_{SEVM} = 0.61$ seconds and $\Delta^{n}_{SEVM} = 5.66$ seconds.
Hence, we have $\beta =  5.66-0.61 = 5.05$.
As $n=4$, we have $\delta_{EVM} = 4 \times 0.61 = 2.44$ so that $\Delta_{EVM} = 5.05+2.44 = 7.49$.
This means that the EVM would take $S = 32\%$ more time than the SEVM to treat these DApp requests.
Finally, as $n$ tends to infinity, the slowdown of the EVM over the SEVM would become $48\%$.

\paragraph{The decoupling of the superblock storage.}

To measure the impact of the decoupling of the reliable storage on the performance of \solution, we
compared the performance of \solution to a naive approach writing the superblock to the disk at once.
To obtain the naive approach, we changed the loop of~\cref{sec:sevm} by simply removing the inner {\ttt for} loop that persisted one sub-block at a time, in order to persist the superblock once and for all. 

\begin{figure}[ht]
	\centering
	\includegraphics[scale=0.27,clip=true,viewport=50 0 930 355]{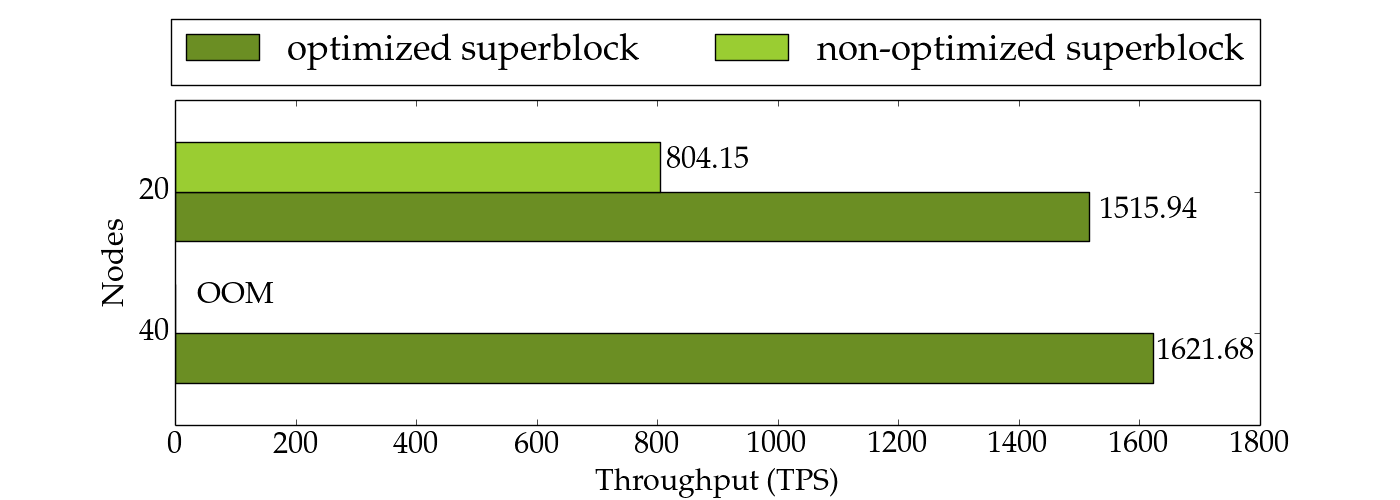}
	\caption{Performance difference when processing each proposal in a superblock (optimized) at a time and when processing the entire superblock in a row (non-optimized) where OOM indicates an out-of-memory error}
	\label{fig:superblock}
\end{figure}

Figure~\ref{fig:superblock} indicates the performance obtained with the SEVM node of \solution (superblock optimized) and with the naive approach (superblock non-optimized). 
At $n=20$ SEVM nodes, the throughput of the superblock optimized version is $1.9\times$ higher than the throughput of the naive version. 
This is because trying to persist a large superblock that comprises $10$ blocks 
leads to IO congestions, stressing the memory 
in order to write to the LevelDB database. 
At $n=40$ SEVM nodes, it is not suprising that the superblock optimized version reaches an even higher throughput than at $n=20$ as \solution scales with the system size. At this scale, however, the superblock non-optimized version crashes by raising an OOM (out-of-memory) error, hence confirming the memory intensive stress of persisting a full superblock at once.

} 

\remove{
\subsection{World-wide deployment with a representative committee}\label{sec:geodistributed}

We now evaluate the performance one can obtain by electing a representative committee of consensus nodes \deepal{blockchain nodes} that decides on the next block. Intuitively, reducing the amount of nodes that need to agree on the next block from potentially all nodes to a committee of nodes, one can reduce the bandwidth consumption and achieve high performance even at world scale.
We decoupled the EVM nodes from the consensus nodes 
and 
ran fewer consensus nodes than EVM nodes. 
We defer the performance evaluation of reconfiguring by replacing one committee by another in \cref{ssec:reconfig}.

In particular, we deploy each consensus node in each of the 12 aforementioned countries (\cref{ssec:world})
and scatter the remaining 60 EVM nodes equally in these countries. 
We also deploy 12 clients, one per region, each sending at a rate of 4500\,TPS.
Although it may seem insufficient to run 12 consensus nodes because a coalition of 4 nodes would be sufficient to reach a disagreement, note that these 12 nodes are proportionally representative. 
As a result, they are all in distinct countries and distinct jurisdictions as when output by the BFT-STV smart contracts.
That is, the risk of coalition is lower than in other blockchains where Proportionality (Def.\ref{def:governance}) is not ensured. As a first example, in EOS the 21 block producers are more likely from the same 4 countries, Cayman Islands, China, Hong Kong and Singapore~\cite{EOS}.
As a second example, NEO runs on top of 7 nodes. As a third example, Ethereum is typically controlled by between 2 and 4 mining pools~\cite{GBE18,EGJ18}. 

\begin{figure}[ht]
	\centering
\includegraphics[scale=0.275,clip=true,viewport=25 0 930 150]{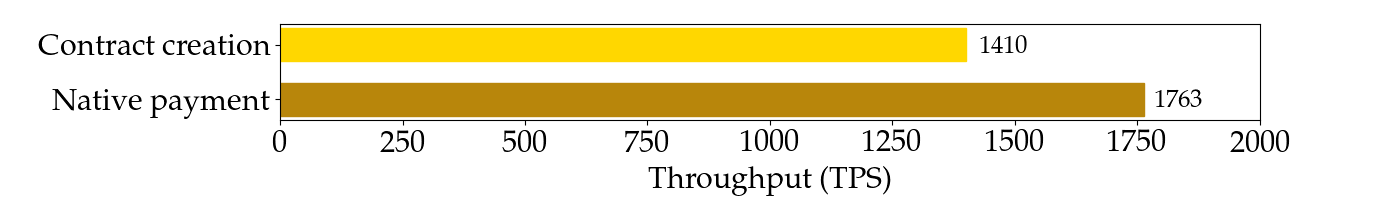}
	\caption{Throughput of \solution when deployed in a geodistributed system of 12 countries across 5 continents\label{fig:geo}}
\end{figure}

Figure~\ref{fig:geo} depicts the throughput of \solution. 
The performance of contract creations and native payments is between 1410 and 1763\,TPS.
The peak throughput is two orders of magnitude faster than Ethereum and one order of magnitude faster than what was reported in the world-wide experiment of SBFT~\cite{GAG19}, which is another blockchain combining an EVM with a leader-based BFT consensus algorithm. This confirms that both the Non-dictatorship design induced by the leaderless property of \solution consensus and helps achieve high performance than other blockchain that require a bottlenecked leader to impose its proposal to the rest of the system.}

\subsection{Reconfiguration performance}\label{ssec:reconfig}
In this section we demonstrate that \solution fully reconfigures its governance with world-wide candidates in less than 5 minutes while ensuring proportional representation of its voters.
More specifically, we evaluate the time it takes to completely reconfigure the blockchain by replacing a committee of blockchain nodes by another (cf. Algorithm~\ref{alg:recon}). To this end, we capture the time it takes to execute the BFT-STV smart contract (i.e., the execution time of BFT-STV) and
start all the newly elected blockchain nodes after initiating a stop on the previous blockchain committee (i.e., the restart time). We define this time as the reconfiguration time (i.e., reconfiguration time =  execution time of BFT-STV + restart time).


\subsubsection{Realistic network delays}

In order to mimic realistic Internet delays in our controlled setting, 
we deployed 20 VMs on OpenStack with 2 VMs per blockchain node, each dedicated for the state service and the consensus service (\cref{sec:lifecycle})\remove{10 consensus nodes and 10 state nodes where}. We added communication delays between the blockchain nodes taken from the AWS geo-distributed environment latencies that we measured separately.
Note that this is 
to 
evaluate the restart time of blockchain nodes in a geo-distributed setting.

\begin{figure}[t]
\centering
\includegraphics[scale=0.27,clip=true,viewport=65 -5 975 400]
{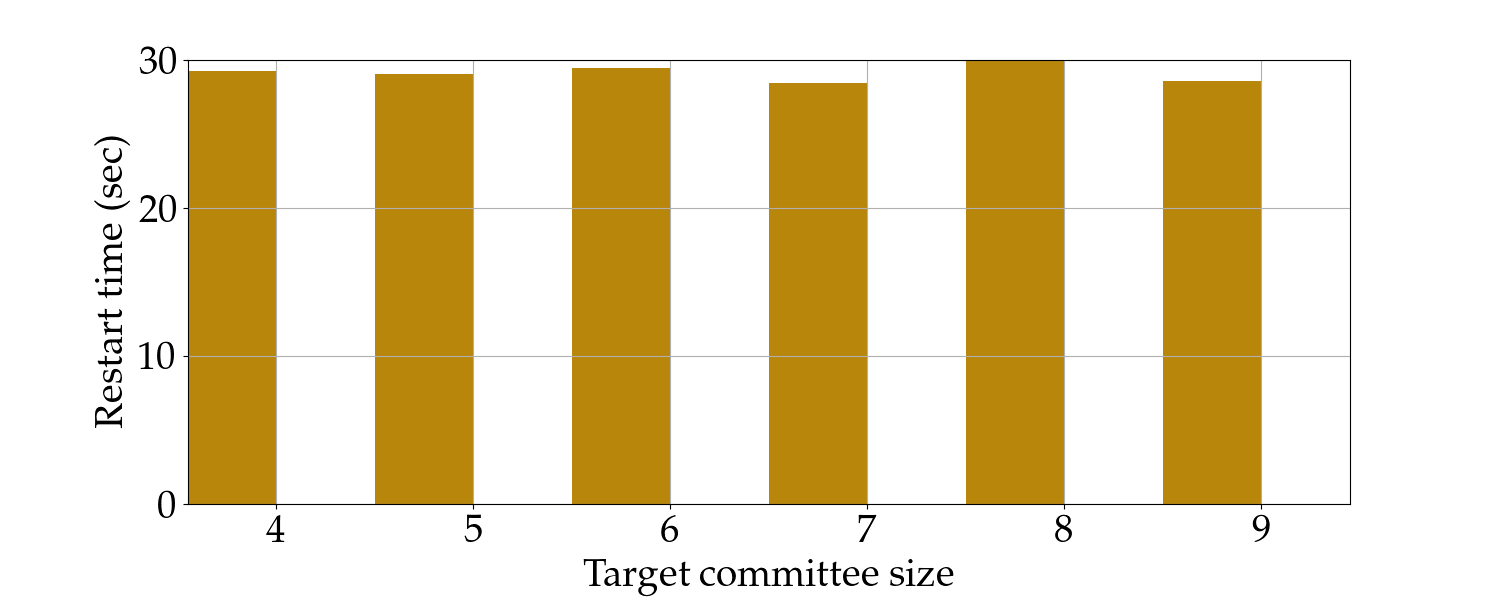}
    \vspace{-2em}
\caption{Restart times in seconds when changing the consensus committee from 10 to target size $k$}
\label{fig:outage}
	\vspace{-1.5em}
\end{figure}

\subsubsection{Restart time}
Figure~\ref{fig:outage} presents the restart time in seconds when a committee of varying size gets started from a total of 10 consensus blockchain nodes.  In particular, we evaluate the restart time when it changes the targeted committee size starting from 10 and targeting a size ranging between 4 and 9 in order to observe whether this size affects the reconfiguration time. 
Note that the lower the time the more available the blockchain is because this time can translate into system outage during which the newly requested transactions may not be serviced---without violating liveness~\cite{GKL15,CS20} (Def.~\ref{def:blockchain}).\footnote{Blockchain liveness remains guaranteed despite reconfiguration as either the correct blockchain node does not receive the transaction due to transient outages or commits it eventually.}
We observe that the restart time varies  by 5\% depending on the targeted committee size but that the maximum time is taken when targeting a committee size of 8 nodes, while the minimum time is taken when targeting a committee size of 7 nodes. This indicates that the restart time is not impacted by the selected targeted committee sizes. To assess the impact of large committee sizes, we experiment below with a larger committee, with 150 voters ranking 150 candidates.

\subsubsection{Impact of the complexity on reconfiguration}\label{ssec:complexity}
Unlike Ethereum or Bitcoin, \solution does not unnecessarily incentivize thousands of participants to execute the same task, hence \solution can restrict the committee size. 
It is well-known that voters of the STV algorithm have to execute an NP-hard computation, so the same applies to the BFT-STV problem. By contrast with concomitant proposals~\cite{CS20b} that approximate other NP-hard proportional election problems, our solution achieves proportionality exactly (Theorem~\ref{theorem1}). Without approximating the solution, our exact solution could induce a cost growing super-linearly with the input size (e.g., number of voters and candidates) of the problem. 
To measure the impact of these parameters, we varied the number $n$ of voters and the number $m$ of candidates from 50 to 150 while executing BFT-STV to reduce the committee of $m/2$ governors. Recall that $m\leq n$ (\cref{ssec:gov}), 
so while we fixed $m=50$ while varying $n$, we had to fix $n=150$ to vary $m$ up to 150. 

\begin{figure}[t]
	\centering
	\includegraphics[scale=0.4]{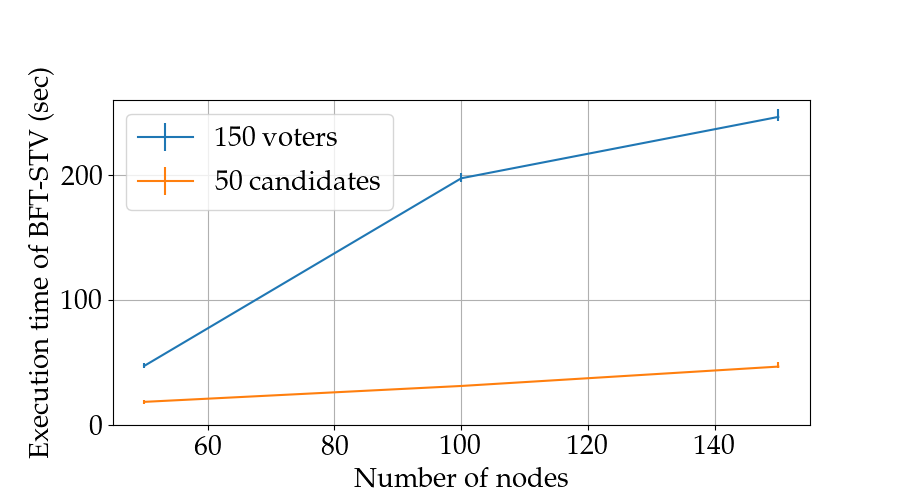}
	\vspace{-2em}
	\caption{The execution time of BFT-STV as we vary the number of candidates (with 150 voters) and as we vary the number of voters (with 50 candidates)}
	\label{fig:STV-exec}
	\vspace{-1em}
\end{figure}

Figure~\ref{fig:STV-exec} depicts the median execution times of BFT-STV and its errors bars (as minima and maxima) over 3 runs for each different pair of numbers $n$ and $m$ of voters and candidates, respectively. 
To this end, for each of these three runs, we generated a random ordinal ballot for each voter.
More precisely, the top curve varies the number $m$ of candidates whereas the bottom curve 
varies the number $n$ of voters. 
\begin{enumerate}
\item First, we observe that the number of candidates impacts significantly the performance with $n=150$ voters, which confirms our expectation. However, we also observe that the raise decreases as $m$ exceeds $100$.
We conjecture that this is due to the way the Ethereum Virtual Machine~\cite{Woo15} garbage collects and alternates between CPU resource usage for transaction execution and I/O usage to persist the information when the transaction increases.
\item 
Second, we observe that when the number of voters increases with $m=50$ the execution time increases sub-linearly: it doubles while the number of voters triples. This is because increasing the number $n$ of voters helps candidates reach the quota $q_B$ of votes rapidly without transferring the vote excess. Hence  the committee is elected faster than expected and raises only slightly the execution time (despite the extra loop iterations due to the increased number of voters). Overall, electing a committee of 75 blockchain nodes in under 4 minutes from $150$ voters ranking $150$ candidates demonstrates the practicality of the BFT-STV smart contract.
\end{enumerate}

Note that ensuring diversity by selecting governors from different countries limit the committee to less than 200 governors anyway as there are only 195 universally recognized nations.
The combination of the results of Figures~\ref{fig:outage} and~\ref{fig:STV-exec} indicates that a \solution governed by each of the OECD members or Asian countries or European countries could reconfigure in less than 5 minutes. Note that this is remarkably short compared to the 3-minute expected time 
taken to execute a single transaction in Ethereum (the Ethereum congestion can make the transaction execution time much longer~\cite{SFG19}).
While it was expected that the execution of the BFT-STV would take longer as we increase the number of candidates and voters (just like the complexity of STV increases with the problem input size), it was unclear whether \solution can scale with the number of machines running the blockchain.
To this end, in the next section (\cref{sec:scalability}), we evaluate the scalability as the impact of the number of VMs participating in the blockchain protocol on its performance.

%

\remove{
Figures~\ref{fig:} and~\ref{fig:} present the performance obtained when automatically replacing the consensus nodes, when executing the representative election of new consensus nodes and when reconnecting the SEVM nodes to the new consensus nodes with a latency between \vincent{XXX and YYY across ZZZ machines from our OpenStack cluster}.
As one can see from Figure~\ref{fig:}, the time it takes to automatically shuffle the consensus nodes and reconnects the SEVM nodes to the newly selected ones is around $30$ seconds and does not change significantly with the size of the targeted committee.}

\subsection{Large-scale evaluation}\label{sec:scalability}
In order to better assess the performance at large-scale, we measure the throughput and latencies when running 100 nodes.
For this experiment we used native payment transactions of 107 bytes (we evaluate \solution later with our BFT-STV DApp).
\remove{\begin{smallitem}
\item {\bf Native payment} corresponds to a simple transaction that transfers a value of 1 wei from one account address to another. 
\item {\bf Smart contract creation} corresponds to a transaction that uploads a smart contract to the blockchain. In particular, we deploy a simple ``Hello, World!'' contract.

\item {\bf Smart contract invocation} corresponds to a transaction that invokes a {\ttt Hello} function of a simple smart contract that returns a byte32 "Hello, World!" message. 
To make sure there is a write operation every time the {\ttt Hello} function is called, we assign ``Hello, World!'' to 
a variable {\ttt message} inside the function.
\end{smallitem}
}

\begin{figure}[t]
	\centering
	\includegraphics[scale=0.25]{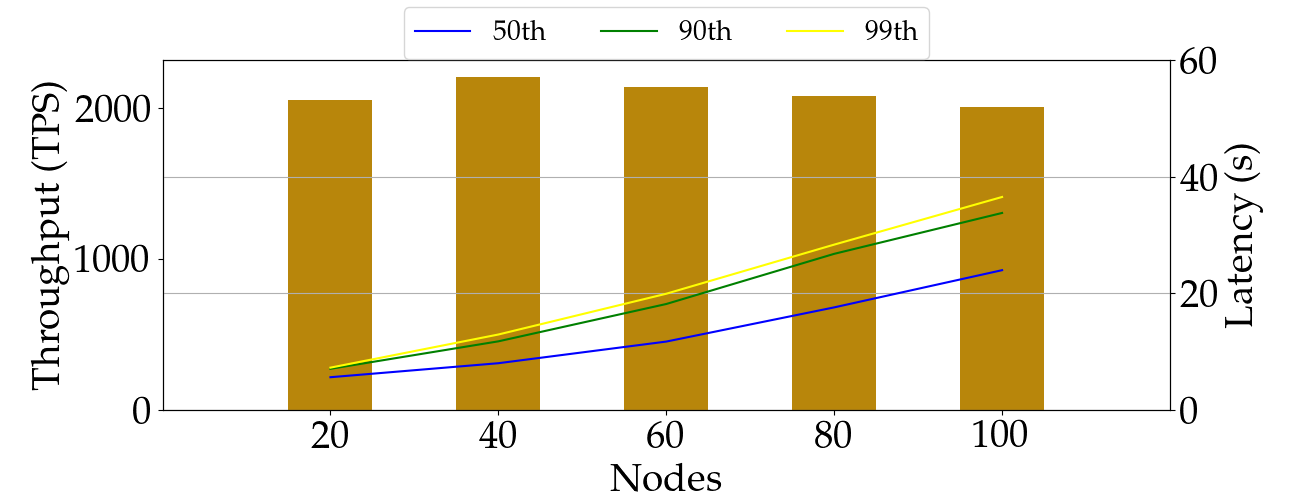}
	\caption{Throughput and latency, expressed as $50^{th}$, $90^{th}$ and $99^{th}$ percentiles, of \solution when executed on from 20 to 100 nodes}
	\label{fig:scal1}
	\vspace{-1em}
\end{figure}

Figure~\ref{fig:scal1} depicts the throughput of \solution with 20 to 100 blockchain nodes in the AWS Sydney availability zone. Each point is the average over 3 runs and clients send 1500 transactions to each of the SEVM nodes, each client sending at a rate of 4500\,TPS.
\remove{First, we observe that the performance varies between 1230 and 1714\,TPS depending on the transaction type.
In particular, the contract creation throughput
is the lowest because each creation writes more data (the whole contract) to the storage trie than an invocation, and a payment only has to update other tries.
By contrast, the native payments and smart contract invocations show higher throughput. Out of these two, the native payments yield the highest throughput as they require the least amount of computation.
Second, we observe that the throughput averaged across all types increases from 1330\,TPS at 20 nodes to 1554\,TPS at 100 nodes due to our validation reduction and superblock optimization.}

We observe that the throughput remains above 2000\,TPS up to 100 nodes and does not changes significantly.
We conclude that \solution performs well in a network of up to 100 nodes. 
These results can be attributed 
to the 
superblock optimization of \solution.

The 50$^{th}$, 90$^{th}$ and 99$^{th}$ percentiles show a steady rise as expected with the increasing number of nodes. This is due to the network overhead in the consensus caused by the increasing number of nodes: 99\% of the transactions are committed within 37 seconds for 100 nodes. 
These latencies are remarkably low compared to what Ethereum already experienced: 22 seconds for the 50$^{th}$ percentile of inclusion time, 2 minutes 39 seconds for the 90$^{th}$ percentile of inclusion time and 47 minutes 34 seconds for the 99$^{th}$ percentile of inclusion time~\cite{SFG19}. 
In addition, an inclusion does not guarantee a commit in Ethereum as a transaction can be included in an uncle block and be discarded and re-included up to 2 more times~\cite{WGP17} later. 
By contrast, the time to commit in \solution remains low
because \solution solves consensus deterministically before appending a new block, hence an inclusion is a commit and 
its user does not have to wait for consecutive appended blocks.





\section{Related Work}\label{sec:rw}

In this section, we present the work related to governance for blockchains.
Table~\ref{table:comparison}, provides a summary of the existing blockchains with reconfigurable governance.
For the sake of brevity, we omit in the discussion below the blockchains that assume synchrony~\cite{Nak08,Woo15,AMN0S17,TR18,8418625,Rchain,BKTFV19} or the ones that were shown vulnerable in the absence of synchrony, including blockchains based on proof-of-authority~\cite{EGJ20}.

\subsection{Proof-of-stake blockchain governance}\label{ssec:pos-rw}
Algorand~\cite{GHM17} was probably the first blockchain assuming that bribery was not instantaneous like we do (\cref{ssec:threat}).
Algorand offers governance through \emph{sortition}, the act of electing governors randomly among a set of candidates.  This technique is similar to the jury selection in trials. 
More precisely, each Algorand node returns a hash $h$ and a proof $\pi$ by passing a publicly known pseudo random input string 
to a verifiable random function locally signed. 
This hash $h$ helps select a node at random 
depending on the amount of assets 
the corresponding user has at stake. 
The selected node can act as a governor and participate in the consensus by piggybacking the sortition proof $\pi$. 
To mitigate bribery attacks, Algorand replaces governors at each step of the BA consensus protocol within a consensus round.
The key advantage of the sortition is that it is a non-interactive cryptographic technique that prevents the adversary from predicting the future governors. 
However, Algorand does not aim at offering any proportional representation.

Polkadot~\cite{cevallos2020verifiably} rotates its governors every \emph{era}, a period that lasts about one day, with a multi-winner election.
Similarly, our BFT-STV is a multi-winner election (\cref{sec:stv}) but can replace governors every five minutes (\cref{ssec:setup}).
Another difference is that Polkadot exploits a nominated proof-of-stake (NPOS):
nominator nodes cast ballots with their preferred candidates based on various parameters (e.g., security practices, staking levels and past performance).
The nominators are rewarded with a fraction of the governors gain upon block creations. The key of NPOS is that the
more stake a candidate has, the more chance it has to be preferred by a candidate and to eventually become a governor. 
A nice advantage over Algorand is that Polkadot's election offers proportional justified representation that limits under-representation.
To avoid overrepresentation and to make it expensive for an adversary to impose its candidates in the election, Polkadot approximates a maxmin objective, but fails at protecting against dictatorship.
%

EOS~\cite{EOS-DPOS} runs a delegated multi-winner approval voting system to elect 21 governors while ensuring some form of proportionality.
As opposed to our BFT-STV (\cref{sec:stv}), approval voting does not allow a voter to indicate its preference between its preferred candidates.
As a result EOS cannot solve our secure governance problem (Def.\ref{ssec:gov}).
EOS exploits delegated proof-of-stake (DPOS) where token holders elect governors by casting a vote whose weight is proportional to what the tokens holders have at stake.
The elected governors run the consensus and as long as voters who own 2/3 of the stake vote for a block proposed by governors, this block is appended to the chain.
EOS may fork in which case the longest chain is adopted as the main one. By contrast, \solution never forks to avoid risky situations where conflicting transactions in different branches lead to double spending.


The aforementioned solutions weight each vote based on the wealth or assets the corresponding voter owns: the more they own the higher weight their vote gets.
Given the Pareto Principle~\cite{Par64} stating that few users typically own most of the resources (as an example in 2021, the wealthiest 1\% of US citizens owned about 1/3 of the total wealth\footnote{\url{https://www.cnbc.com/2021/10/18/the-wealthiest-10percent-of-americans-own-a-record-89percent-of-all-us-stocks.html}.}), these approaches are vulnerable as soon as one manages to bribe the few wealthiest of all nodes as they likely control a large part of the total stake. 

\subsection{Proof-of-work blockchain governance}

Zilliqa~\cite{ZILLIQA} is a sharded blockchain that supports smart contracts and reaches consensus using an efficient version of the leader-based consensus protocol PBFT~\cite{CL02} based on EC-Schnorr multisignature~\cite{syta2016keeping,kogias2016enhancing}.
To shard the network and to reach consensus on transactions, a committee of directory service (DS) nodes is elected
with a proof-of-work 
(PoW) puzzle.
Once a candidate node finds the nonce for the PoW, it generates a DS header and multicasts a DS block to the current DS committee. Once the current DS committee reaches consensus on the DS block mined and multicast by the candidate node, the new candidate node is added to the DS committee and the oldest member of the DS committee is removed. The protocol thus ensures that the latest $n$ nodes that have mined a DS block
are governors.

The hybrid consensus~\cite{pass2017hybrid} is a theoretical consensus algorithm for blockchain that selects the most recent $\ell$ block miners as governors.
Similar to Zilliqa, each governor is replaced one at a time following a leader-based consensus algorithm. 
Unfortunately, we are not aware of any implementation of the hybrid consensus. 

Both approaches
need as many leader-based consensus executions as there are governors to completely rotate them. 
By contrast, \solution rotates them all in a single consensus instance to mitigate the risks of an adversary bribing progressively most of the current governors.
\subsection{BFT blockchain governance}

The vast majority of byzantine fault tolerant (BFT) blockchains assume that the list of governors is selected by an external service.
ComChain~\cite{VG19} lists the public keys of governors in configuration blocks but assumes that the new lists of governors are proposed by an external service. Similarly, Tendermint/Cosmos~\cite{TC21} lists the public keys of governors in blocks 
but 
associates a voting power to each validator based on its stake, hence risking the same bribery attacks as other proof-of-stake blockchains (\cref{ssec:pos-rw}).
SmartChain~\cite{BAS20} also stores the committee public keys in dedicated reconfiguration blocks but simply grants governor credentials to every requesting node, without requesting to go through an election. 
Libra~\cite{BBC19} mentions a similar reconfiguration service but no details are provided regarding the selection of governors.
As far as we know other BFT blockchains have a static set of consensus nodes, which makes them more vulnerable to bribery attacks, including Stella~\cite{LLM19}, SBFT~\cite{GAG19}, Concord~\cite{concord} and Quorum~\cite{jpmorganchase_quorum}.

\remove{

\subsection{Payment blockchains}
Some blockchains are designed for high transactions throughput at large scale, but were not designed for DApps~\cite{GHM17,CNG18,SDV19,LLM19,GRH20}. This is the case of ResilientDB~\cite{GRH20} that exploits topology-awareness to parallelize consensus executions, the
Red Belly Blockchain~\cite{CNG18} that shares our superblock optimization or Mir~\cite{SDV19} that 
 deduplicates transaction verifications.
Stellar~\cite{LLM19} is an in-production blockchain running in a geodistributed setting.

\subsection{Towards Byzantine fault tolerant blockchains}
Upper-bounding the number $f$ of Byzantine failures allow to solve consensus to avoid forks.
Ethereum comes with proof-of-work and 
proof-of-authority (PoA) in the two mainstream Ethereum programs, called {\ttt parity} and {\ttt geth}. The idea of proof-of-authority is to have a set of $n$ permissioned validators, among which $f$ can be malicious or \emph{Byzantine}~\cite{PSL80}, that generate new blocks~\cite{noauthor_ethereum_nodate,noauthor_poacore_nodate,BBK18}.
Unfortunately, 
both proof-of-authority protocols in {\ttt parity} and {\ttt geth} have recently been shown vulnerable to the attack of the clone when messages take longer than expected~\cite{EGJ20}. 

\subsection{Tolerating unpredictable bounded delays}
To cope with unpredictable message delays, blockchains cannot rely on synchrony.
Ethermint~\cite{Ethermint} is a blockchain that combines the partially synchronous Tendermint consensus protocol~\cite{BKM18} with the EVM. At the time of writing, Ethermint is in its ``pre-alpha development stage'' and could not be evaluated.
In particular, we found some issues that prevented us from deploying it, like a lack of support for hexadecimal addresses~\cite{Eth20} and a nonce management limitation, which resulted in rejecting consecutive transactions sent in a short period of time~\cite{Eth19}. Other researchers who managed to deploy an older version of it, reported a peak throughput of 100\,TPS obtained with a single validator node~\cite{DBD18}.

Zilliqa~\cite{ZILLIQA} is a blockchain that supports smart contracts and reaches consensus with PBFT~\cite{CL02}. 
We are not aware of any performance evaluation of Zilliqa but its  state machine, Scilla, executes non Turing complete programs slower than the EVM when the state size increases~\cite{Scilla}.
Therefore, it is unlikely that it would yield higher throughputs than our \solution for large state sizes.
Chainspace~\cite{albassam2017chainspace} introduced a distributed atomic commit protocol termed S-BAC for  smart contract transactions. Coupled with the BFT-SMaRt~\cite{BSA14} consensus protocol, Chainspace can support trustless use of DApps. However, 
it has only been able to achieve up to 350\,TPS, offering a limited support for DApps.

\subsection{Evaluations of BFT blockchains}
Quorum~\cite{jpmorganchase_quorum} is a blockchain that supports Ethereum smart contracts and reaches consensus with the Istanbul Byzantine Fault Tolerant (IBFT) consensus algorithm. 
Just like \solution, the Byzantine fault tolerance of Quorum makes it well-suited for mobile devices to interact wth DApps securely without downloading the blockchain. Moreover, it seems that few optimizations could help it treat a large number of transactions per second~\cite{BSK18}. Unfortunately, our performance evaluation of Quorum indicated that in its current form, Quorum cannot sustain high demands and loses requests as the workload increases, which confirms recent observations~\cite{SNG20}. 


SBFT~\cite{GAG19} is a Byzantine fault tolerant consensus algorithm that exploits threshold signatures to reduce the communication complexity of PBFT but commits, like PBFT, at most one proposed 
block per consensus instance. It was shown to reach consensus on 378 smart contract requests per second when deployed within one continent and 172 requests per second across multiple continents. 
Concord~\cite{concord} is a blockchain that combines a lightweight C++ implementation of the EVM with SBFT, however, its publicly available version has open issues~\cite{bugconcord20} that prevent it from being deployed on distinct physical machines but we showed that Concord, although slower than \solution, reached the encouraging throughput of 1000\,TPS on 4 nodes within the same physical machine. It could be the case that future versions will scale. 


\subsection{Sharding techniques}
Sharding~\cite{Eth2,FBP20,PRISM} is known
to improve the performance of blockchains. 
%

The {\ttt move} approach~\cite{FBP20} moves accounts and computation from one smart contract enabled blockchain to another. The smart contract of the first blockchain is locked before any participant creates it in the second blockchain. This allows to scale the throughput of the congested DApp CryptoKitties with the number of shards.
Eth2~\cite{Eth2} relies on a beacon chain and will feature 64 shard chains to improve the scalability of Ethereum. The uniqueness of the beacon chain guarantees a consistent view of current state but cannot handle accounts and smart contracts. The validators of the shard chain do not need to download and 
run data for the entire network.

{\sc Prism}~\cite{PRISM} is a proof-of-work blockchain that shards the blockchain into $m$ voter chains and exploits three types of blocks in a block tree.
The voter blocks are used to vote for proposer blocks grouped per level in the block tree.
Once a proposer block is elected, transaction blocks that are pointed to by the proposer block are committed. 
Prism peaks at 19K\,TPS by 
ignoring the eager validation completely,
which exposes it to DoS attacks.
%
To ensure the copy of the blockchain state is not corrupted, a user needs first to download the block headers, a time- and space-consuming task ill-suited for running DApps on handheld devices.

Sharding is orthogonal to our approach. Exploiting multiple instances of \solution in as many shards would likely increase its performance, but as opposed to the solutions above, this would not need a DApp to download any shard chain.

} 


\section{Conclusion}\label{sec:conclusion}

We presented \solution, a blockchain that solves the swift proportional governance problem by electing a governance committee that is proportionally representative of the voters and by reconfiguring itself fast with this new governance.
Its novelty lies in tolerating $f<n/3$ byzantine governors, preventing the adversary from acting as a dictator and 
reconfiguring sufficiently fast to cope with bribery attacks.
Our evaluation shows that \solution is practical and performs efficiently at 100\,nodes.
This research bridges the gap between computational social choice and blockchain governance and opens up new research directions
related to bribery mitigation and non-dictatorship.
%
%

\bibliographystyle{ACM-Reference-Format}
\bibliography{reference} 

\appendix 
\section{Proofs of Blockchain Safety, Liveness and Validity}
\label{line:proof}
In this section, we show that \solution solves the  blockchain problem (Def.~\ref{def:blockchain}). For the sake of simplicity in the proofs, we assume that there are as many nodes playing the roles of consensus nodes and state nodes and are collocated on the same physical machine. 

\begin{lemma}
	\label{theorem1}
	If at least one correct node $\lit{propose}$s to a consensus instance $i$, then 
	every correct node decides on the same superblock at consensus instance $i$.
\end{lemma}

\begin{proof}
	The $\lit{propose}(\cdot)$ function is the same as in the Red Belly Blockchain~\cite{CNG21} except that we do not use the verifiable reliable broadcast and reconciliation which ensure SBC-validity. Instead we use reliable broadcast and no reconciliation. As such, following from the proof of~\cite{CNG21}, our blockchain nodes ensure SBC-termination that states that every correct node eventually decides on a set of transactions and SBC-agreement that states no two correct nodes decide on different sets of transactions. Since \solution consensus returns a superblock (which is a set of transactions) at each instance of consensus $i$, we can say every correct blockchain node decides on the same superblock at consensus instance $i$.
\end{proof}

\begin{lemma}\label{lemma:safety}
	Variables $TX_k$, $R_k$, $S_{next_{k}}$, $\ms{timestamp}_{k}$ and $val_{k}$ become identical after executing line~\ref{line:assign-rk} of iteration $k$ for any two correct blockchain nodes $P1$ and $P2$.
\end{lemma}
\begin{proof}
	From Lemma~\ref{theorem1}, every correct consensus node decides on the same superblock at consensus instance $i$. As a result, each correct blockchain node receives the same superblock from consensus instance $i$ (Algorithm~\ref{alg:sevmexecution}, line~\ref{line:execution}). Therefore, each $\ms{props[k]}$ for any integer $k \in [0;n)$ becomes identical at any two correct blockchain nodes $P1$ and $P2$ (Algorithm~\ref{alg:sevmexecution}, line~\ref{line:props}) for consensus instance $i$. As a result, $TX_k$, $R_k$, $S_{next_{k}}$, $timestamp_{k}$ and $val_{k}$ also become identical at $k$ for $P1$ and $P2$ for consensus instance $i$ (Algorithm~\ref{alg:sevmexecution}, lines \ref{line:start-var}--\ref{line:end-var}).
\end{proof}
\begin{lemma}\label{lemma:safety2}
	At each index $\ell$ of the chain,  all correct blockchain nodes can only append the same block 
	$B_\ell$.
	
\end{lemma}

For the next lemma we refer to $B_0$ as the genesis block of \solution.

\begin{proof}
	The proof is by induction on the index of the blocks in the chain.
	\begin{itemize}
		\item {\bf Base case:}
		if $\ell=1$, then $h(H(B_{0})) \in B_{1}$.
		Since $h(H(B_{0}))$ is the hash of the header of the genesis block, it is the same for all correct nodes of \solution. 
		We know from Lemma~\ref{lemma:safety} that variables $h(S_{next_{k}}), h(\ms{TX}_k), h(R_k), \ms{val}_{k}$ and $\ms{timestamp}_{k}$ are identical after  line~\ref{line:assign-rk} onwards for any two correct blockchain nodes $P1$ and $P2$. Therefore, ${B_{1}}$ is identical for all correct blockchain nodes.
		\item {\bf Inductive case:}
		Let us assume that $B_{\ell-1}$ is identical for all correct blockchain nodes, we show that 
		$B_{\ell}$ is identical for all correct blockchain nodes.
		Since  $B_{\ell-1}$ is identical, $h(H(B_{\ell-1}))$ is identical for all correct blockchain nodes. Using the argument used in the base case, any variable $h(S_{next_{k}})$, $h(TX_k)$, $h(R_k)$, $val_{k}$ and $timestamp_{k}$ are identical after line~\ref{line:assign-rk} onwards for any two correct blockchain nodes $P1$ and $P2$.
		Since
		$B_{\ell} \gets h(H(B_{\ell-1})), h(\ms{S_{next_{k}}}), h(TX_{k}), h(R_{k}), GU_{k}, GL$, $nonce$, $timestamp_{k}, val_{k}$, it must also become identical for all blockchain nodes after  line~\ref{line:assign-rk} onwards.
	\end{itemize}
	Therefore, by induction on each index $\ell$, $B_{\ell}$ becomes identical for all correct blockchain nodes when $B_\ell$ is constructed at line~\ref{line:end-var}.
\end{proof}

The next three theorems show that \solution satisfies each of the three properties of the blockchain problem (Definition~\ref{def:blockchain}).
\begin{theorem}
	\solution satisfies the safety property.
\end{theorem}
\begin{proof}
	The proof follows from the fact that any block $B_{\ell}$ at index $\ell$ of the chain is identical for all correct blockchain nodes due to Lemma~\ref{lemma:safety2}.
	
	Due to network asynchrony, it could be that a correct node $P1$ is aware of block $B_{\ell+1}$  at index $\ell+1$, whereas another correct node $P2$ has not created this block $B_{\ell+1}$ yet.
	At this time, $P2$ maintains a chain of blocks that is a prefix of the chain maintained by $P1$.
	And more generally, the two chains of blocks maintained locally by two correct blockchain nodes are either identical or one is a prefix of the other. 
\end{proof}

\begin{theorem}
	\solution satisfies the validity property.
\end{theorem}	
\begin{proof}
	From Algorithm~\ref{alg:sevmexecution},	line \ref{line:valid2}, only valid transactions are executed and added only such valid transactions are added to block $B_{\ell}$ (Algorithm~\ref{alg:sevmexecution} - lines \ref{line:validity1} and \ref{line:end-var}). Therefore, $\forall$ indexes $\ell$, $B_{\ell}$ is valid $\forall$ correct blockchain nodes.
\end{proof}

\begin{theorem}\label{thm:liveness}
	\solution satisfies the liveness property.
\end{theorem}	
\begin{proof}
	As long as a correct replica receives a transaction, we know that the transaction is eventually proposed by line~\ref{line:waittimer} of Algorithm~\ref{alg:selection}. 
	The proof follows from the termination of the consensus algorithm and the fact that \solution keeps spawning new consensus instance as long as correct replicas have pending transactions.
	The consensus algorithm is DBFT~\cite{CGLR18} and was shown terminating with parameterized model checking~\cite{BGK21}.
\end{proof}

\section{Discussion}\label{sec:discussion}

Although \solution offers the swift governance reconfiguration by solving the secure governance problem (Def.~\ref{def:governance}) and the blockchain problem (Def.~\ref{def:blockchain}), there are few secondary aspects that need to be detailed or addressed. Below, we discuss these aspects and propose 
different extensions to \solution.

\paragraph{Know-your-customer (KYC) initial selection}
In \solution, we employ a know-your-customer (KYC) process when deploying \solution for the first time in order to assign candidate and voter permissions to blockchain nodes. 
Similar to VeChain~\cite{vec} or the EOS-based Voice social network\footnote{\url{https://crypto-economy.com/eos-based-social-network-voice-announces-human-sign-up-an-alternative-for-kyc/}},
we require potential voters and candidates to provide personal information to a decentralized foundation before they can be granted the desired permissions. 
This personal information can include
the name of the user, their biometric details and their preferred role and can be traded against a permission to act as a candidate for election or a voter. The foundation then verifies each user personal information and, when the verification is successful, assigns the specified role to the corresponding user and a one-time secret sent through a secure channel to join \solution. In \solution, each voter node is given equal opportunity to elect a candidate to the committee of blockchain nodes. 
This KYC-based solution thus copes with the risks of building an oligarchy of wealthiest users through proof-of-stake (PoS) or of most powerful machines through proof-of-work (PoW).
Although this KYC-based solution requires an offchain verification that is external to the blockchain, note that it is only needed at bootstrap time: any blockchain node deciding to run the code of \solution (including the BFT-STV smart contract codes) necessarily accepts that the set of voters can automatically elect new candidates and new voters every $x$ blocks without the control of any external company, jurisdiction or foundation.

A possible attack is to upload a bribing smart contract that rewards voters if a specific set of candidates
is elected.
We underscore that this cannot happen in-band, since such a malicious smart contract needs to be deployed for such a process to take place, 
and since we assume less than $1/3$ of nodes are byzantine at all times, such deployments are impossible as a majority would not agree when reaching consensus.

\paragraph{Domain Name Service (DNS)}
For the sake of simplicity in the design of our solution, we implicitly assumed that the IP addresses were static so that nodes would simply need to subscribe to the BFT-STV smart contract to receive an emit event informing them of a list of IP addresses (Algorithm~\ref{alg:stv} line~\ref{line:emit}) and allowing thems to reconnect to the blockchain nodes running the consensus service.
Although this assumption is suitable when experimenting on a controlled set of VMs provided by a cloud provider,
 the IP addresses of Internet users are often dynamically assigned, which makes our solution unrealistic. 
One can easily adapt the current implementation to support domain names instead of IP addresses. While this solution requires a Domain Name Service (DNS) that, if centralized, could defeat the purpose of the blockchain, note that the hard-coded DNS server addresses are already used by classic blockchains for node discovery~\cite{Nak08}.
\solution can exploit DNS in a similar fashion but to offer governance reconfiguration by updating promptly the DNS so as to redirect all clients to the new committee and to mitigate long-range attacks.
	
\paragraph{Gas cost of reconfiguration execution}
As \solution builds upon the Ethereum Virtual Machine (EVM)~\cite{Woo15}, it inherits the notion of gas that avoids the infinite execution of smart contracts.
The gas is a unit that reflects the work required for a particular sequence of execution steps in Ethereum~\cite{Woo15}. The more computation is required to execute a transaction, the more gas this transaction requires to complete. It is the client's role to pay each unit of gas that is consumed by the transaction it issued in the blockchain. The price of this gas unit is known as the gas price and is considered as the transaction fee to reward the blockchain node (called `miner' in Ethereum) who mined the block containing the transaction. In the BFT-STV smart contract, the election execution cost between 340029258 and 2347870086 gas units. The gas price depends on the network traffic and if a client requires to have their transactions treated as priority by miners, then they should include a high gas price. However, if we use a high gas price, the transaction cost becomes extremely high and almost unaffordable for a voter.
To mitigate this issue, we allow transactions that invoke the $\lit{cast-ballot}$ function at line~\ref{line:castballot} of Algorithm~\ref{alg:stv} to be treated with equal priority and have priority over the rest of the transactions despite using a low gas price. This allows, voters to participate in the election without incurring high costs. 


\paragraph{Period of committee change}
As we already mentioned in Figure~\ref{fig:bftstv}, \solution starts a new configuration every $x$ blocks, where $x=100$ by default.
However, as \solution produces block on-demand (after receiving sufficiently many transactions), triggering the next reconfiguration could take a very long time, if for example transactions were issued rarely.  
Instead, we need to rapidly reconfigure \solution so that $n$ voters get re-elected before $n/3$ of them get corrupted through a bribery attack. Otherwise, an adversary could gain progressively the control of a coalition of $n/3$ or more voters to finally dictate the decisions to the rest of the blockchain nodes.
To cope with this issue, we require every blockchain node offering the \solution service to spawn {\ttt no-op} transactions on a regular basis (line~\ref{line:waittimer} of Algorithm~\ref{alg:selection}). As blockchain nodes
get rewarded based on the service they offer, this reward can be used to compensate the loss associated by these {\ttt no-op} transaction fees.
Provided that the clock skews between correct blockchain nodes is bounded, then these {\ttt no-op} transactions should ensure that reconfiguration occurs sufficiently frequently to cope with bribery attacks.



\paragraph{Privacy, pseudonymity and anonymity}\label{app:anonymity}
To maintain the integrity of the voting protocol and to prevent users from influencing one another, the voting protocol should remain anonymous~\cite{CCC20}.
As the smart contract is stored in cleartext in the blockchain data structures, a honest but curious adversary could easily map the public key of a voter to its cast ballot. This preserves pseudonymity: as long as this user does not link publicly its identity to its public key, then it remains anonymous. 
To offer strong anonymity, \solution can be combined with
a commit-reveal scheme~\cite{kirillov2019implementation,zhao2015vote} or ring signature~\cite{CCCG20} to reveal the voter once the election terminates, or homorphic encryption~\cite{10.1145/3459087} or zk-SNARK~\cite{tarasov2017internet} on top of \solution. 

\onecolumn

\section{BFT-STV Smart Contract}\label{app:sc}
{\tiny
\begin{lstlisting}[language=Solidity,basicstyle=\scriptsize\LSTfont,escapechar = ?]
pragma solidity $^0.4.0$;
pragma experimental ABIEncoderV2;

contract Committee {
	
	string [] committee;
	address public chairperson;
	mapping(string => bool) hasIp;
	mapping(string => bool) hasCalled;
	mapping(address => string) WallettoIP;
	mapping(uint => string []) ballots; // mapping between ballot number and the vote transferrable to the next round
	mapping(string => uint) votes;
	mapping(bytes32 => uint) rest;
	mapping(uint => uint) ballot-index;
	string [] public selected;
	string [] surplus-current;
	uint32 [] private digits;
	bytes32 [] hashes;
	uint member;
	mapping(string => uint) transfer-vote;
	mapping(string => uint) tot-surplus;
	mapping(uint => uint) indexers;
	uint c-ballot;
	uint threshold;
	uint size;
	uint select;
	uint transfer;
	uint k;
	uint rest-tot;
	uint quota;
	uint min;
	uint eliminatedcount;
	uint round;
	bool val;
	bool excess;
	bool eliminated;
	bool elect;
	mapping(string => bool) elected;
	event notify(string []);
	constructor() public {
		chairperson = msg.sender;
		c-ballot = 0;
	}

	// initial set of node ips and the size of the committe is parse by the chairperson
	function addIp (string [] memory ip, uint members, string[] memory wallets) public {
		delete committee;
		require(
			msg.sender == chairperson,
			"Only chairperson can give right to vote."
		);
		// committee here is the initial set of nodes -- this is equal to candidates
		for (uint t = 0; t < ip.length; t++){
			committee.push(ip[t]);
			WallettoIP[parseAddr(wallets[t])] = ip[t];
			elected[ip[t]] = false;
			hasIp[ip[t]] = true;
			hasCalled[ip[t]] = false;
		}
		size = committee.length;
		member = members;
		threshold = size - (size - 1)/3;
		quota=(threshold/(member+1))+1;
		// members is the number of participants per committee
	}
	
	
	// you have to change this function according to STV
	function createCommittee (string [] memory candidates) public {
		// if the caller of this function is in the list of ips added by the chairperson, and if they haven't call this function
		// before - because we don't want t+1 be reached by a malicious node calling this function multiple times
		require(hasIp[WallettoIP[msg.sender]] == true & hasCalled[WallettoIP[msg.sender]] == false);
		hasCalled[WallettoIP[msg.sender]] = true;
		
		// check if the IPs of the candidates received are the same as the addIP candidates
		for(uint i=0; i<candidates.length; i++) {
			for(uint z=0; z<committee.length; z++) {
				if(keccak256(abi.encodePacked(candidates[i]))==keccak256(abi.encodePacked(committee[z]))){    
					ballots[c-ballot].push(candidates[i]);
					votes[candidates[i]]=0;
					transfer-vote[candidates[i]]=0;
				}
			}
		}
		if(ballots[c-ballot].length !=0) {
			c-ballot=c-ballot + 1;
		}
		if(c-ballot == threshold) {
			// start doing stv
			for(uint a=0; a<c-ballot; a++) {
				ballot-index[a]=0;
			}
			round = 0;
			eliminatedcount=0;
			// first preference vote calculation
			for (a = 0; a<c-ballot; a++){
				votes[ballots[a][round]]=votes[ballots[a][round]]+1;
			}
			
			// do until the number of seats-members is filled
			while ((selected.length < member) & round<size) {
				// loop through ballot and count votes
				elect = false;
				excess = false;
				eliminated = false;
				// add changes from here //
				val=next-pref();
				if (val==true) {
					elect = true;
				}
				if (elect == false) {
					// remove least voted
					min = 100000;
					// minimum vote of all candidates (non elected) is eliminated
					for (a = 0; a<c-ballot; a++) {
						for(uint x=ballot-index[c-ballot]; x<size; x++) {
							if((votes[ballots[a][x]] < min) & !elected[ballots[a][x]]){
								min = votes[ballots[a][x]];
							}
						}
					}
					eliminate();
				}
				if ((size-eliminatedcount)==member) {
					break;
				}
				round = round + 1;
			}
			
			
			// to add the remainder of members //
			min = minimum();
			while((size-eliminatedcount) >member) { 
				for (a = 0; a<c-ballot; a++) {
					for(x=0; x<size; x++) {
						if(votes[ballots[a][x]]==min & (size-eliminatedcount) >member & votes[ballots[a][x]]!=10000000){
							votes[ballots[a][x]]=10000000;
							eliminatedcount=eliminatedcount+1;
							min = minimum();
						}
					}
				}
			}
			// add the new section here - if the seats to be selected equals the non eliminated candidates (won and not yet elected),
			// assign already non elected and non eliminated candidates to the seats and exit the loop.
			if((size - eliminatedcount) == member) {
				for(a = 0; a<c-ballot; a++) {
					for(x=0; x<size; x++) {
						if(votes[ballots[a][x]]!=10000000 & !elected[ballots[a][x]] & (selected.length)<member){
							selected.push(ballots[a][x]);
							elected[ballots[a][x]]=true;
						}
					}
				}
			}
		
		emit notify(selected);
	}
	
	
	function minimum() returns (uint mins){
		mins = 100000;
		// minimum vote of all candidates is eliminated
		for (uint a = 0; a<c-ballot; a++){
			for(uint x=0; x<size; x++){
				if((votes[ballots[a][x]] < mins) & !elected[ballots[a][x]]){
					mins = votes[ballots[a][x]];
				}
			}
		}
		return mins;
	}
	
	
	function next-pref() public returns (bool elec){
		for (uint s=0; s<committee.length; s++) {
			transfer-vote[committee[s]]=0;
			tot-surplus[committee[s]]=0;
		}
		for (s=0; s<hashes.length; s++) {
			rest[hashes[s]] = 0;
		}
		surplus-current.length=0;
		elect=false;
		excess=false;
		
		if (selected.length < member) {
			
			for (s=0; s<c-ballot; s++) {
				for (uint x=ballot-index[s]; x<size; x++) {
					if (votes[ballots[s][x]]>=quota & votes[ballots[s][x]]!=10000000 & elected[ballots[s][x]] == false) {
						elected[ballots[s][x]] = true;
						selected.push(ballots[s][x]);
						elect = true;
						transfer = votes[ballots[s][x]] - quota;
						votes[ballots[s][x]]=quota;
						transfer-vote[ballots[s][x]] = transfer;
						surplus-current.push(ballots[s][x]);
						ballot-index[c-ballot]=x;
						excess = true;
					}
					
				}
			} 
			
			if (excess) {
				// break the surplus ballots to those that have the same surplus elected candidate
				for(uint a=0; a< c-ballot; a++) {
					indexers[a] = 7000;
				}
				for (uint j=0; j<c-ballot; j++) {
					for (k=0; k<surplus-current.length; k++) {
						if (keccak256(abi.encodePacked(ballots[j][ballot-index[j]]))==keccak256(abi.encodePacked(surplus-current[k]))) {
							tot-surplus[ballots[j][ballot-index[j]]] = tot-surplus[ballots[j][ballot-index[j]]] + 1;
							x = 1;
							if ((ballot-index[j]+x)<(size -1)) {
								while((elected[ballots[j][ballot-index[j]+x]] || votes[ballots[j][ballot-index[j]+x]]==10000000) 
								& (ballot-index[j]+x) < (size -2)) {
									x = x + 1;
								}
								
								if(!elected[ballots[j][ballot-index[j]+x]] & (ballot-index[j]+x)<size){
								    rest[keccak256(abi.encodePacked(ballots[j][ballot-index[j]],ballots[j][ballot-index[j]+x]))] = 	rest[keccak256(abi.encodePacked(ballots[j][ballot-index[j]],ballots[j][ballot-index[j]+x]))] + 1;	hashes.push(keccak256(abi.encodePacked(ballots[j][ballot-index[j]],ballots[j][ballot-index[j]+x])));
									// mapping for ballot number => current indexer
									indexers[j]=ballot-index[j]+x;
								}
							}
						}
					}
				}
				// divide and add the transfered votes
				for(j=0; j< c-ballot; j ++) {
					for(x=1; x<(size-1); x++) {
						//maybe this might throw an error-look into it later
						if ((ballot-index[j]+x) < size & rest[keccak256(abi.encodePacked(ballots[j][ballot-index[j]],
							ballots[j][ballot-index[j]+x]))] != 0 
							& !elected[ballots[j][ballot-index[j]+x]] 
							& tot-surplus[ballots[j][ballot-index[j]]]!=0  
							& votes[ballots[j][ballot-index[j]]] != 10000000 
							& votes[ballots[j][ballot-index[j]+x]]!=10000000) {
								votes[ballots[j][ballot-index[j]+x]] = votes[ballots[j][ballot-index[j]+x]]
									+ transfer-vote[ballots[j][ballot-index[j]]] 
									* rest[keccak256(abi.encodePacked(ballots[j][ballot-index[j]],ballots[j][ballot-index[j]+x]))]
									/ tot-surplus[ballots[j][ballot-index[j]]];
								rest[keccak256(abi.encodePacked(ballots[j][ballot-index[j]],ballots[j][ballot-index[j]+x]))] = 0;         
						}
					}
				}
				// moving the indexer to next //
				for(a=0; a< c-ballot; a++){
					if(indexers[a]!=7000){
						ballot-index[a]=indexers[a];
					}
				}
			}
			
		}
		return elect;
	}
	
	
	function eliminate() {
		uint x;
		for (uint s=0; s<hashes.length; s++) {
			rest[hashes[s]] = 0;
		}
		for (uint a = 0; a<c-ballot; a++) {
			for (uint m=ballot-index[a]; m<size; m++) {
				if (votes[ballots[a][m]] == min & !elected[ballots[a][m]] & votes[ballots[a][m]] != 10000000 & !eliminated) {
					// if 0 just eliminate
					if (votes[ballots[a][m]] == 0) {
						votes[ballots[a][m]] = 10000000;
						eliminated=true;
						eliminatedcount=eliminatedcount+1;
						//if it is the current index being eliminated
						x=1;
						if ((m == ballot-index[a]) & (m<size-1)) {
							while((elected[ballots[a][m+x]] || votes[ballots[a][m+x]]==10000000) & (m+x)<(size -1)){
								x=x+1;
							}
							if (!elected[ballots[a][m+x]] & votes[ballots[a][m+x]]!=10000000) {
								ballot-index[a]=m+x;
							}
						}
						break;
					}
					// otherwise //
					x = 1;
					if (m<(size-1)) { 
						while((elected[ballots[a][m+x]] || votes[ballots[a][m+x]]==10000000) & (m+x)<(size -1)) {
							x = x + 1;
						}
						if (!elected[ballots[a][m+x]] & (m+x)<size) { 
							rest[keccak256(abi.encodePacked(ballots[a][m],ballots[a][m+x]))] = 
								rest[keccak256(abi.encodePacked(ballots[a][m],ballots[a][m+x]))] + 1;
							hashes.push(keccak256(abi.encodePacked(ballots[a][m],ballots[a][m+x])));
						}
					}
				}
			}
		} 
		// transfer to the next preferences
		for (a = 0; a<c-ballot; a++) {
			x=0;
			while(x<(size-1) & !eliminated){
				if (rest[keccak256(abi.encodePacked(ballots[a][x],ballots[a][x+1]))] != 0 
					& votes[ballots[a][x]]!=10000000 
					& votes[ballots[a][x+1]]!=10000000 
					& !elected[ballots[a][x]]) {
					if (votes[ballots[a][x]]!=0) {
						votes[ballots[a][x+1]] = votes[ballots[a][x+1]] 
							+ min*rest[keccak256(abi.encodePacked(ballots[a][x],ballots[a][x+1]))]
							/ (votes[ballots[a][x]]);
						ballot-index[a]=x+1;
					}
					rest[keccak256(abi.encodePacked(ballots[a][x],ballots[a][x+1]))] = 0;
					votes[ballots[a][x]] = 10000000;
					eliminatedcount=eliminatedcount+1;
					eliminated = true;
				}
				x=x+1;
			}
		}
	}
}
\end{lstlisting}
} 

\end{document}